\documentclass[3p]{elsarticle}
\usepackage[utf8]{inputenc}
\usepackage[T1]{fontenc}
\usepackage{amsmath}
\usepackage{amssymb}
\usepackage{graphicx}
\usepackage{fixmath}

\begin{document}

\title{A pair polarimeter for multi-GeV $\gamma$-rays}
\author[a]{Maxime Defurne}
\author[a]{Arthur Muhulet}
\affiliation[a]{organization={Irfu, CEA, Université Paris-Saclay}, postcode={91191}, city={Gif-sur-Yvette}, country={France}}

\begin{abstract}
   Accessing the polarization of photon allows to understand the mechanisms behind its emission or scattering, revealing much about a peculiar environment or a probed object. For energy above $\sim$10~MeV, the pair production dominates the photon-matter interaction and the photon polarization is accessible via the azimuthal angle of the conversion. Unfortunately pair polarimeters have a low figure-of-merit for multi-GeV photons and are mostly used for beam characterization. In this paper, we report a new concept of a compact pair polarimeter associating monolithic active pixel sensors to low-density extended solid converters to reach simultaneously a high efficiency of $\sim$7\% and intrinsic analyzing power ranging from 0.2 to 0.5. This new concept will add a new obersvable to the multi-messenger physics, isolate the intrinsic strong force in nucleons and possibly reveal violations of Lorentz invariance.    
\end{abstract}


\maketitle

\section{Introduction}
The polarization of multi-GeV photon is an experimental observable holding many promises in multiple fields. In Astrophysics, such high-energy $\gamma$-rays are emitted by supermassive blackholes, pulsars or heavy-mass binary systems. Accessing their polarization would reveal much about the environment and structure of these astrophysical objects~\cite{ilie}. The structure of much smaller objects such as hadrons would greatly benefit from the knowledge of the $\gamma$-rays polarization. Indeed deeply virtual Compton scattering is a process in which a virtual photon inelastically scatters off a quark confined in the proton. The polarization of the final state photon is expected to be a key observable to study linearly polarized gluons in the proton, considered as the intrinsic strong field to form the hadron. Finally the structure of vacuum itself can be probed as well by studying the polarization of multi-GeV photons. Exposed to an intense field being either electromagnetic, gravitational or even strong such as in a quark-gluon Plasma, vacuum may become birefringent which would break Lorentz invariance. The birefringence can be highlighted by the alteration of the photon polarization when passing through the vacuum~\cite{ilie,QuantumGravity,Birefringence}. Unfortunately such sources or physics phenomena have low brightness or cross sections. Consequently, to measure the associated linear polarization of a multi-GeV photon with a reasonable statistical accuracy, the polarimeter must have a high Figure-of-Merit.\\

\subsection{Principle of multi-GeV $\gamma$-polarimetry}
 For multi-GeV photons, polarimeters rely on pair conversion as it dominates by much the total cross section of the photon interaction with the matter. The polarization fraction is extracted by studying the distribution of conversions with respect to their azimuthal angle $\phi$. The conversion rate is given by:
\begin{equation}
  \frac{d\Gamma}{d\phi}\propto\left(1+A_cP \cos\left[2\left(\phi-\phi_0\right)\right]\right)\;,
  \label{eq:ConvEq}
\end{equation}
with $A_c$ the analyzing power of the conversion, $P$ the degree of linear polarization and $\phi_0$ the polarization direction. The analyzing power depends on the imbalance in energy between the leptons: it reaches 0.245 when both photons inherits half the photon energy and almost linearly drops to 0 when the photon energy is transfered to a single lepton. Integrated over the energy imbalance, the convolution of both analyzing power and pair production cross sections gives an average $A_c$ of 0.14.\\

The challenge of $\gamma$-polarimetry lies in measuring the azimuthal angle and thus for two reasons. The first reason is the smallness of the opening angle between the two leptons. For a high-energy photon, the most likely opening angle between the two leptons $\theta_{+-}$ is given by:
\begin{equation}
\theta_{+-}\sim \frac{1.6~\textrm{MeV}}{E_{\gamma}}\;,
\end{equation}
where $E_{\gamma}$ is the photon energy expressed in MeV. From 2 to 8~GeV, $\theta_{+-}$ varies between 2$\times 10^{-4}$ and 8$\times 10^{-4}$ rad. With such a low opening angle, distinguishing the two leptons requires space and/or accurate detectors.\\

The second limitation arises from multiple Coulomb scattering blurring the conversion angles. This multiple scattering is an intrinsic feature of the pair polarimeter since the photon must interact with the matter to convert into a lepton pair. The pair polarimeter allows to retrieve the linear polarization of $N_{\gamma}$-photons by studying the azimuthal distribution of the induced conversions given by:
\begin{equation}
\frac{dN_{e^+e^-}}{d\phi}\propto N_{\gamma} \epsilon \left(1+A_c A_{_{I}} P \cos\left[2\left(\phi-\phi_0\right)\right]\right)\;.
\end{equation}

Two additional parameters are introduced compared to Equation~\ref{eq:ConvEq} which are $\epsilon$ the conversion efficiency and $A_{_{I}}$ the intrinsic analyzing power of the polarimeter depending on the detector resolution and the multiple scattering occuring in the polarimeter. Here a trade-off must be found for the radiation length of the converter: the conversion efficiency $\epsilon$ increases with the radiation length whereas the intrinsic analyzing power $A_{_{I}}$ decreases due to multiple scattering. From these parameters, the Figure-of-merit for such a polarimeter is defined by:
\begin{equation}
 \mathcal{F}^2_{\gamma}=A^2_{_{I}} \times A^2_c \times \epsilon
\end{equation}
with:
\begin{equation}
  \epsilon=1-\exp\left(-\frac{7}{9}\frac{X}{X_{0}}\right)\;. 
\end{equation}

Like any polarimeter, the statistical accuracy on the polarization measurement is given by:
\begin{equation}
  \sigma_{P}=\frac{1}{\mathcal{F}} \sqrt{\frac{2}{N_{\gamma}}}
  \label{eq:Stat}
\end{equation}

\subsection{State-of-the-art in $\gamma$-polarimetry via pair production}
\begin{figure*}[!htp]
  \begin{tabular}{cc}
    \includegraphics[width=0.65\linewidth]{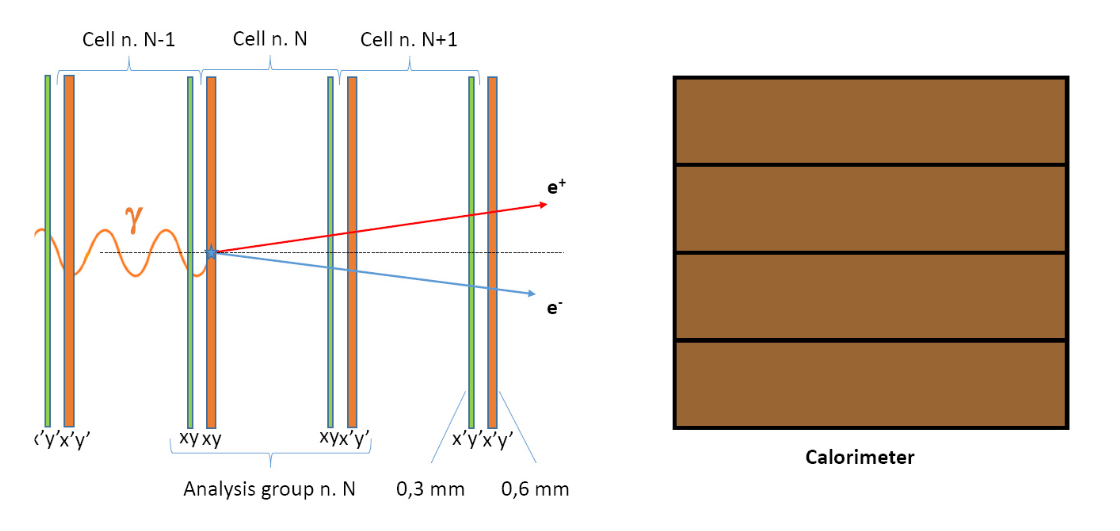}&\includegraphics[width=0.35\linewidth]{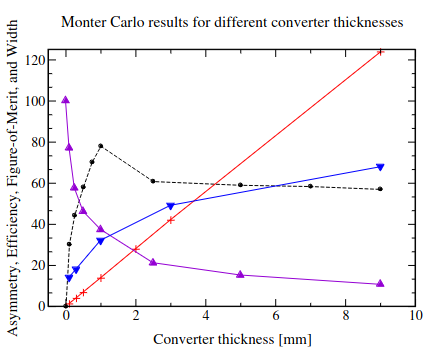}\\
  \end{tabular}
  \caption{\label{fig:bogdan}Left: Schematic of the pair polarimeter. Right: Optimization of the Figure-of-merit (black line) with the converter thickness. The red line is the conversion rate, the purple line the analyzing power of the polarimeter $A_{_{I}}$ and the blue line the resolution on the azimuthal angle. Figures are from~\cite{bogdan}.}
\end{figure*}

At \emph{low} energy, 10-20~MeV, a Time Projection Chamber (TPC) offers a resolution good enough to resolve the lepton pair: The photon converts in the gas and the lepton pair is seen by the ionization of the latter from the conversion vertex. A prototype of such a concept named HARPO~\cite{HARPO} was successfully built and caracterized by CEA, CNRS and \'Ecole Polytechnique. However the maximal photon energy for polarimetry is limited by the TPC spatial resolution and the conversion efficiency is low as the converter is a gas. Now there are pair polarimeters with very thin solid converters (0.2\% $X_{0}$) and trackers located downstream of a dipole to resolve the lepton pair~\cite{de_Jager_2004}. However the size and the low conversion efficiency makes it unable to study polarization of at most one million uncollimated photons. The most relevant design of a pair polarimeter for high-energy $\gamma$-rays is presented in Eingorn \emph{et al.}~\cite{bogdan} and shown in Figure~\ref{fig:bogdan}.
The polarimeter consists of 30 cells, each made of a two double-sided micro-strip detectors (MSD) with a pitch of 50~$\mu$m and separated by 20~mm, a first thick MSD acting as converter and a second thin MSD as analyzer (to determine the azimuthal angle of the lepton pair). For 500~MeV-gamma rays, the cell length must be 6.5~cm to resolve the two leptons with an accuracy good enough to measure the azimuthal angle. The radiation thickness of the converter MSD was optimized: while the effective analyzing power of a cell decreases with the thickness due to multiple scattering, the conversion efficiency increases faster until one percent of radiation length where a maximum in the Figure-of-Merit is reached as seen in Figure~\ref{fig:bogdan}.\\

\section{A revisited concept of pair polarimeter for DVCS}
\label{ssec:pair}
The design of polarimeter proposed by Eignorn \emph{et al.} has greatly inspired the subsequent work. For multi-GeV photons, a first limitation comes from the spatial resolution of MSDs which is not good enough unless the length of the cell is dramatically increased. Here we suggest to use monolithic active pixel silicon sensors (MAPS). With pixels being 25$\times$25 $\mu$m$^2$, MAPS detectors achieve 5$\mu$m-resolution. An example of such detector is the Muon Forward Tracker of ALICE consisting of ALPIDE sensors~\cite{MAGER2016434}. Another significant advantage is the thinness of the silicon layer being only 50 $\mu$m with a detection efficiency above 99\%. Including the support layers (Kapton, Alumimium, Carbon,...) the material budget in the active region is approximately 0.25\% X/X$_0$. Nowadays an intense R\&D effort is still ongoing to reduce even more the material budget of MAPS as well as their power consumption. However the low radiation length of MAPS decreases the efficiency of the polarimeter. To compensate this loss without increasing the number of MAPS layer and consequently the cost of the apparatus, an independent converter must be added to the cell design. 

\begin{figure*}[!htp]
  \centering
  \begin{tabular}{cc}
    \includegraphics[width=0.65\linewidth]{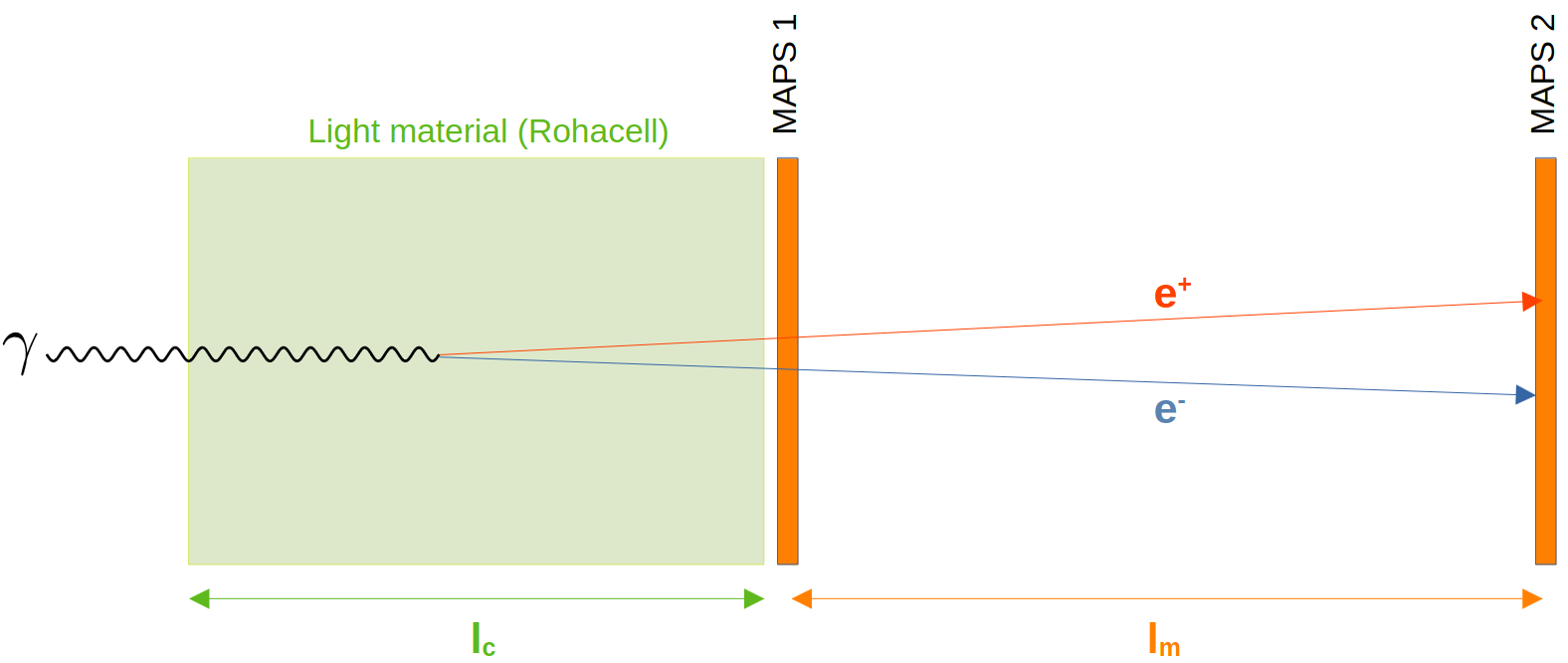}& \includegraphics[width=0.35\linewidth]{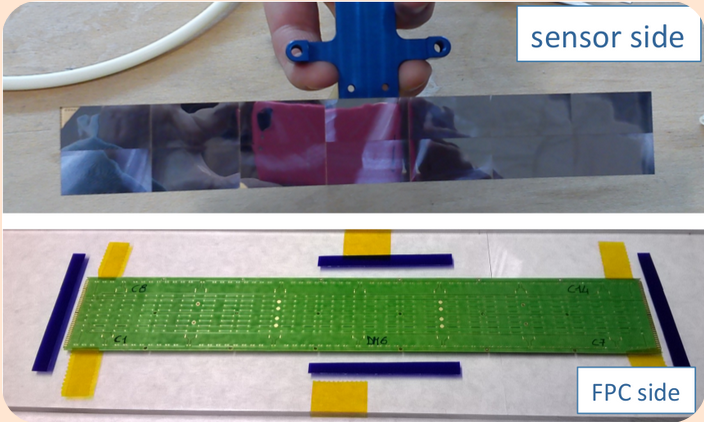}
  \end{tabular}
  \caption{\label{fig:GammaPol} Left: Schematics of the cell proposed for GluToN$\gamma$. Right: ALPIDE MAPS detectors used by ALICE muon forward tracker with a 5$\mu m$-resolution.}
\end{figure*}

Looking back to Eingorn's design, a second limitation stems from the thinness of the converter: All the multiple scattering happens in the vicinity of the conversion vertex and the information about the conversion angles are immediately/maximally blurred. Here we suggest to fill the cell with a low-density material as displayed by Figure~\ref{fig:GammaPol}. Unlike a thin converter, the multiple scattering in a low-density material will happen all along the track instead of immediately after the conversion. As the information is recovered from the position measurements in the MAPS, a deviation of the particle is much less significant when it occurs close to the measurement planes than at the conversion vertex. The extended converter could be some high-pressure heavy gas or low-density material such as Rohacell 31/51/71.\\

In the following, we are going to first describe the modelisation of the MAPS response to Minimum Ionizing Particles (MIPS) using ALPIDE sensors as reference. Then tracking algorithms and optimization studies will quantify all the effects previously mentionned, leading to an optimized pair polarimeter design. 

\subsection{A realistic MAPS description} \label{sec:MAPS}
First, the single-particle response will be optimized to reproduce the ALPIDE characteristics presented in~\cite{MAGER2016434}. Then a study of cluster shapes caused by two close particles going through the MAPS plane is performed to extract as much information as possible to be used at a later stage in a Kalman filter for the polarimeter study. 

\subsubsection{MAPS response to single particle}
When a MIP goes through a thin silicon sensor such as an ALPIDE sensor, it deposits some energy following a Landau-Vavilov distribution by ionization. The Most Probable Value (MPV) of this distribution ($\Delta_{_p}$) is:
\begin{equation}
  \Delta_{_p} \simeq \xi \left[\ln{\frac{2mc^2\xi}{(\hbar\omega_{_p})^2}}+0.200\right]\;,
\end{equation}
where the definitions of the plasma energy $\hbar\omega_{_p}$ and the parameter $\xi$ can be found in the section entitled \emph{passage of particles through matter} in the PDG~\cite{PDG}. The width of the distribution is $w \simeq 4\xi$. Following this distribution, an energy deposit is randomly generated for each particle in the Toy Monte-Carlo. This energy is then converted into a number of ionized electron/hole pairs placed uniformly along the path of the particle through the fully depleted zone.\\

\begin{figure*}[!htp]
    \includegraphics[width=0.9\linewidth]{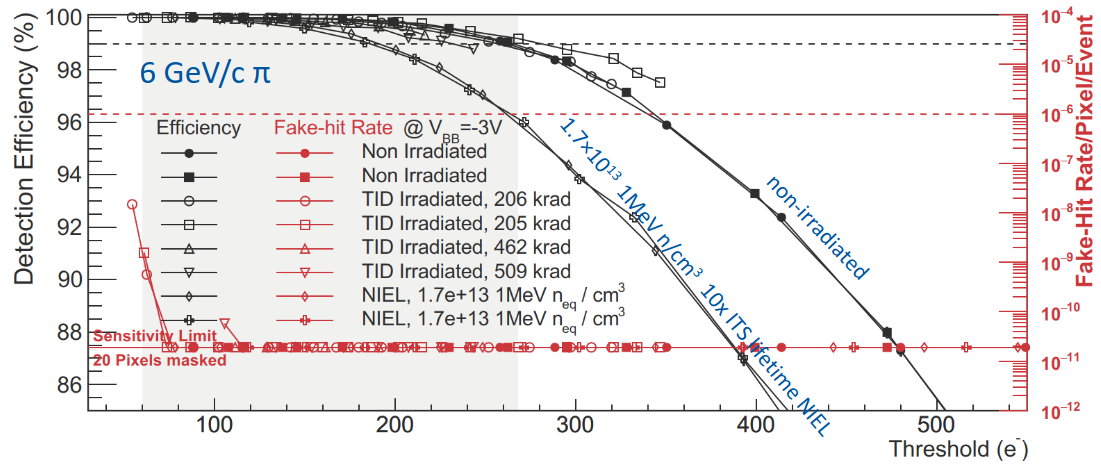}
    \includegraphics[width=0.9\linewidth]{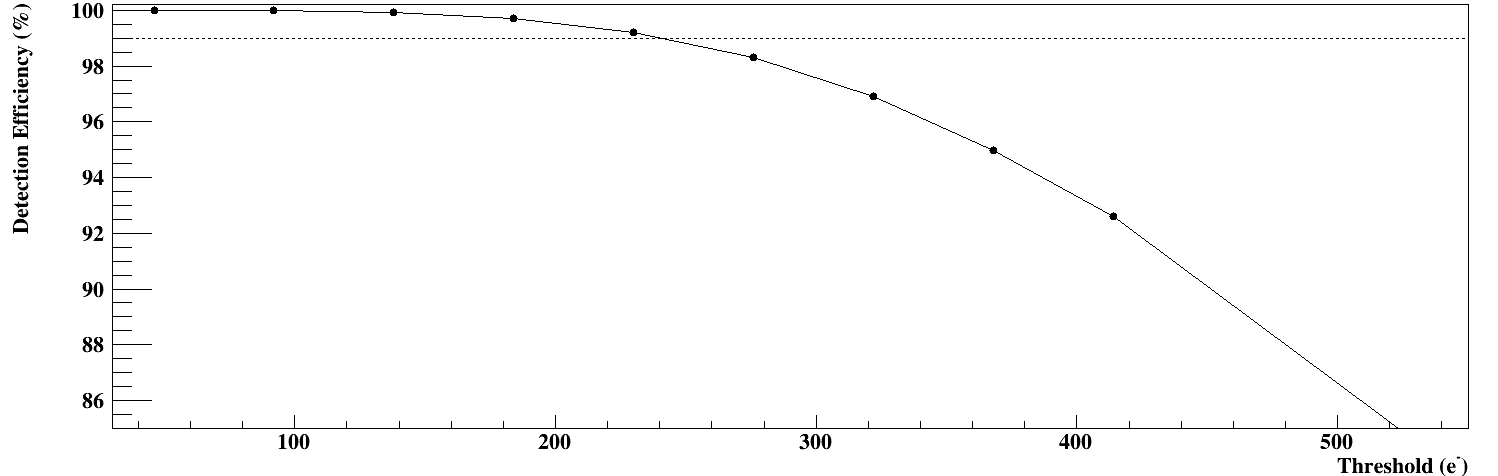}
    \includegraphics[width=0.9\linewidth]{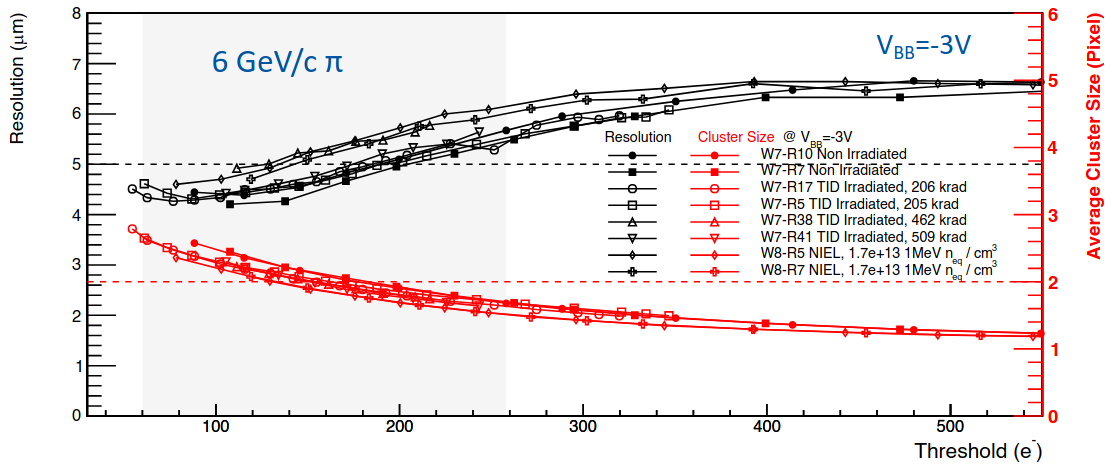}
    \includegraphics[width=0.9\linewidth]{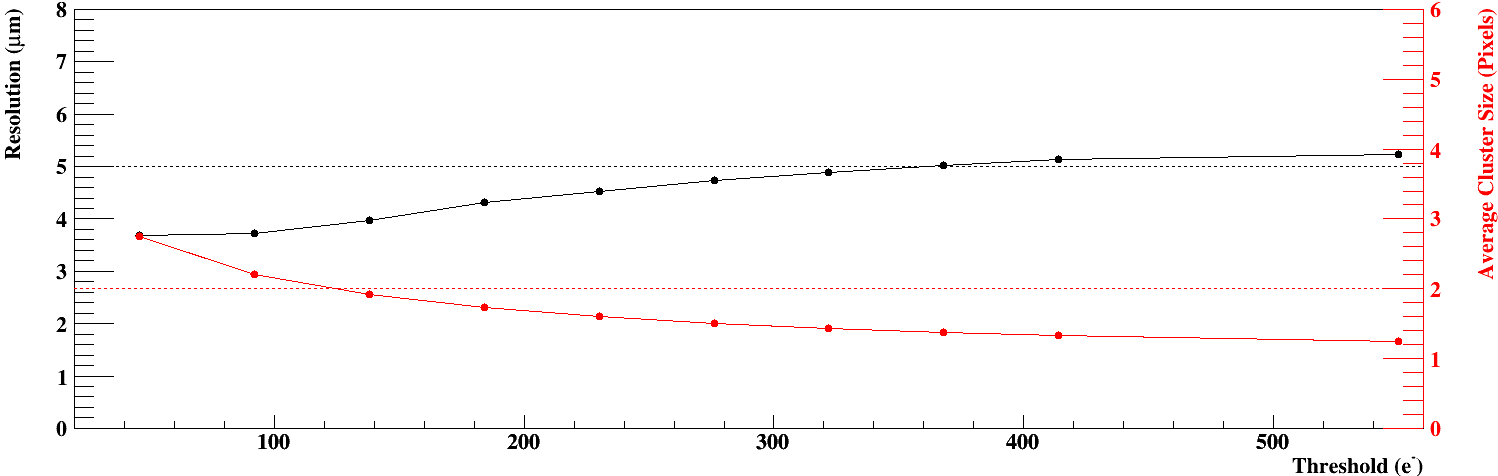}
  \caption{\label{fig:sizeMAPS} Top: Efficiency as function of the threshold for a true~\cite{MAGER2016434} (top) and a toy-MC (bottom) ALPIDE sensor. Bottom: Resolutions and average cluster size as function of the threshold for a true~\cite{MAGER2016434} (top) and a toy-MC (bottom) ALPIDE sensor.}
\end{figure*}

Their distance to the collection diode, in the middle of the pixel, is calculated as well as the average electric field seen on the path. The average electric field is converted into an average velocity of the hole $v_{_p}$ with the following relation:
\begin{equation}
  v_{_p}=\mu_{_p}\vec{E}\;,
\end{equation}
where $\mu_{_p}$ is the mobility of a hole. The hole mobility depends on the bias voltage and the doping density but here is taken as a constant value of 150 ~cm$^{2}$/Vs. From the velocity and the distance, the drift time $t_{_d}$ is derived. The longitudinal and transverse diffusion of holes is then given by:
\begin{equation}
\sigma=\sqrt{2Dt_{_d}}\;,
\end{equation}
where $D$ is the hole diffusion constant ($D$=10.86 ~cm$^{2}$/s). In order to reproduce as best as possible ALPIDE caracteristics presented in the article~\cite{MAGER2016434}, the MPV and width $w$ of the Landau-Vavilov distribution had to be multiplied by 0.28: For very thin absorber such as 50$\mu$m-thick silicon sensors, the MPV is known to be smaller than the results from the PDG formula. Displayed by Figure~\ref{fig:sizeMAPS}, both efficiency and cluster size distributions as function of the threshold are faithfully reproduced whereas the resolution is found 1$\mu$m better than the measured ones.

\begin{figure}[!htp]
  \centering
    \includegraphics[width=0.575\linewidth]{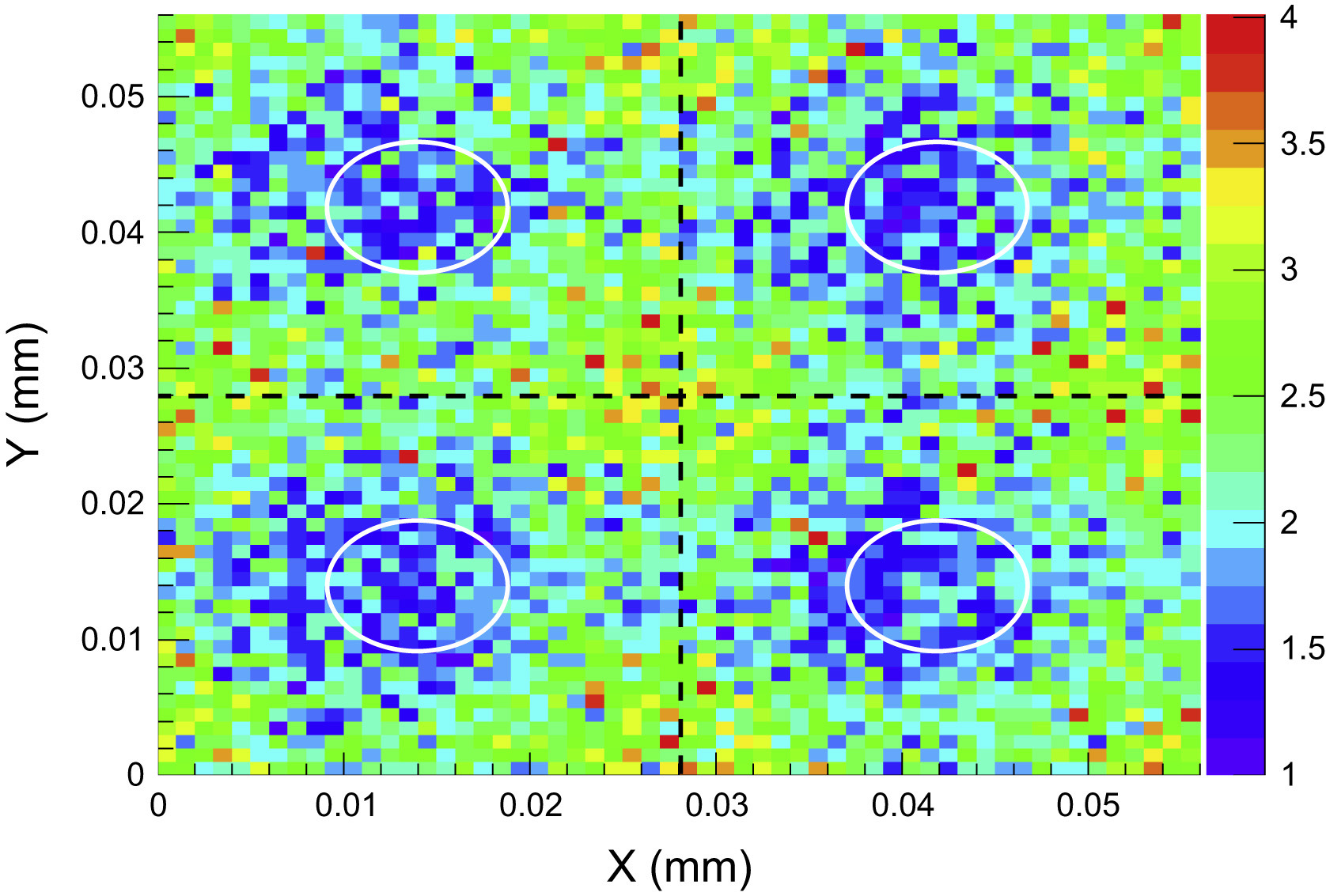}
    \includegraphics[width=0.395\linewidth]{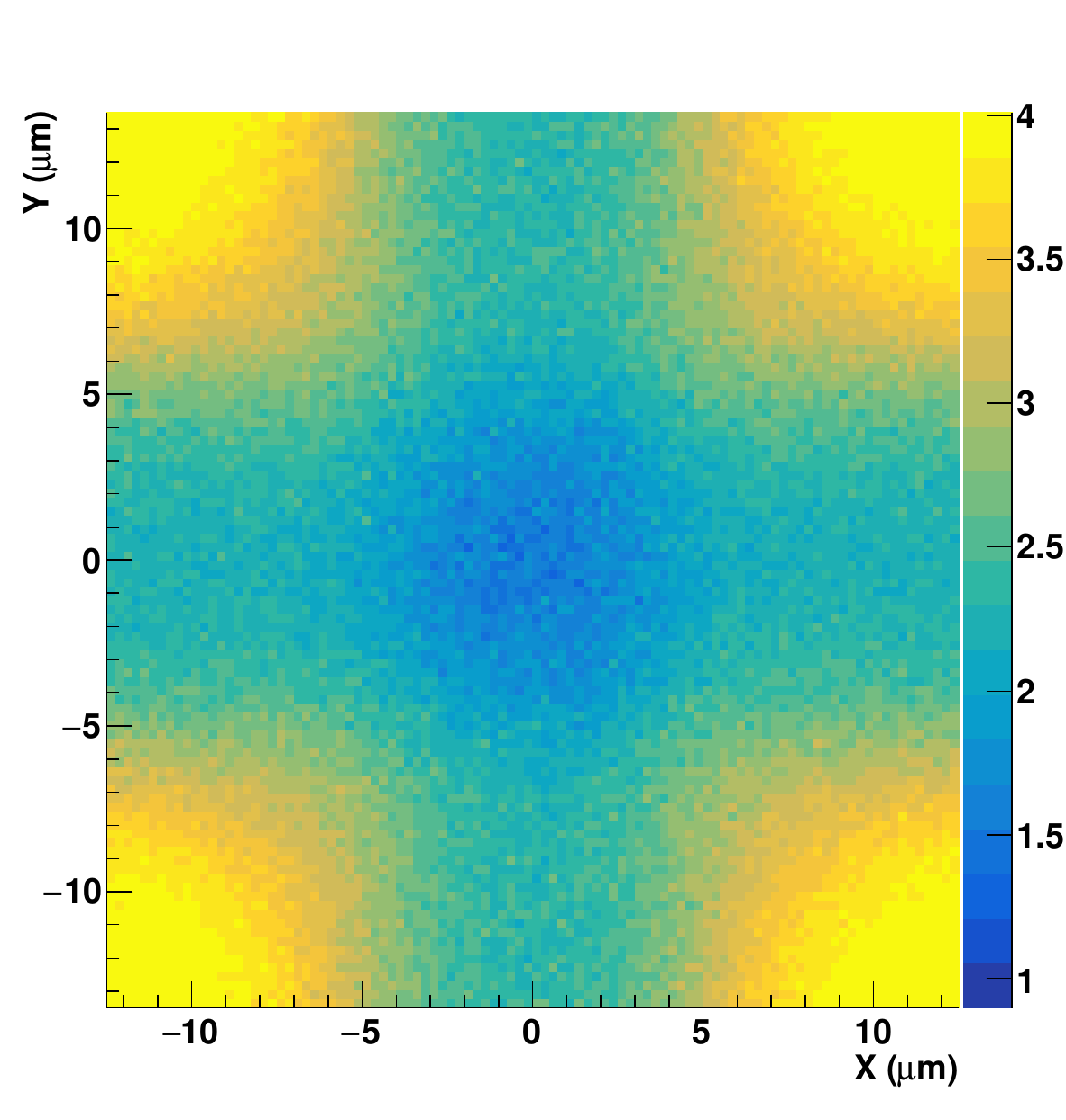}
  \caption{\label{fig:cluster2D} Average cluster size as function of the particle position within a pixel for a true~\cite{MAGER2016434} (left) and a toy-MC (right) ALPIDE sensor.}
\end{figure}

In addition to these fundamental properties, the average cluster size as function of the particle location in a pixel are compared in Figure~\ref{fig:cluster2D}. When the particle hits the middle of a pixel, the average cluster size is close to 1. As the particle gets closer to a neighbor pixel either on the same row or column, the average cluster size increases to 2. When getting closer to two neighbor pixels by the diagonale of the pixel, L-shaped 3-pixel clusters become the main pattern which may turn into a squared 4-pixel cluster when the particle hits the corner of a pixel. \\

This geometrical analysis of a single-particle cluster shape can be extended to 2-particle single-cluster event. Using this toy MC model, probability maps for 2-particle positions are inferred for each single-cluster or close-two-cluster patterns.

\subsubsection{MAPS response to two close particles}
First, a single particle is shot at a random position in the middle pixel of a 9$\times$9-pixel matrix. Then, at a random distance of at most 100 $\mu$m from the first particle, a second particle is generated. The threshold is set to 90 electrons to consider a pixel being fired. The matrix is then analyzed to determine the number of clusters and their relative position resulting from the convolution between energy deposit and the particle positions. \\

\begin{figure}[!htp]
  \centering
  \includegraphics[width=0.66\linewidth]{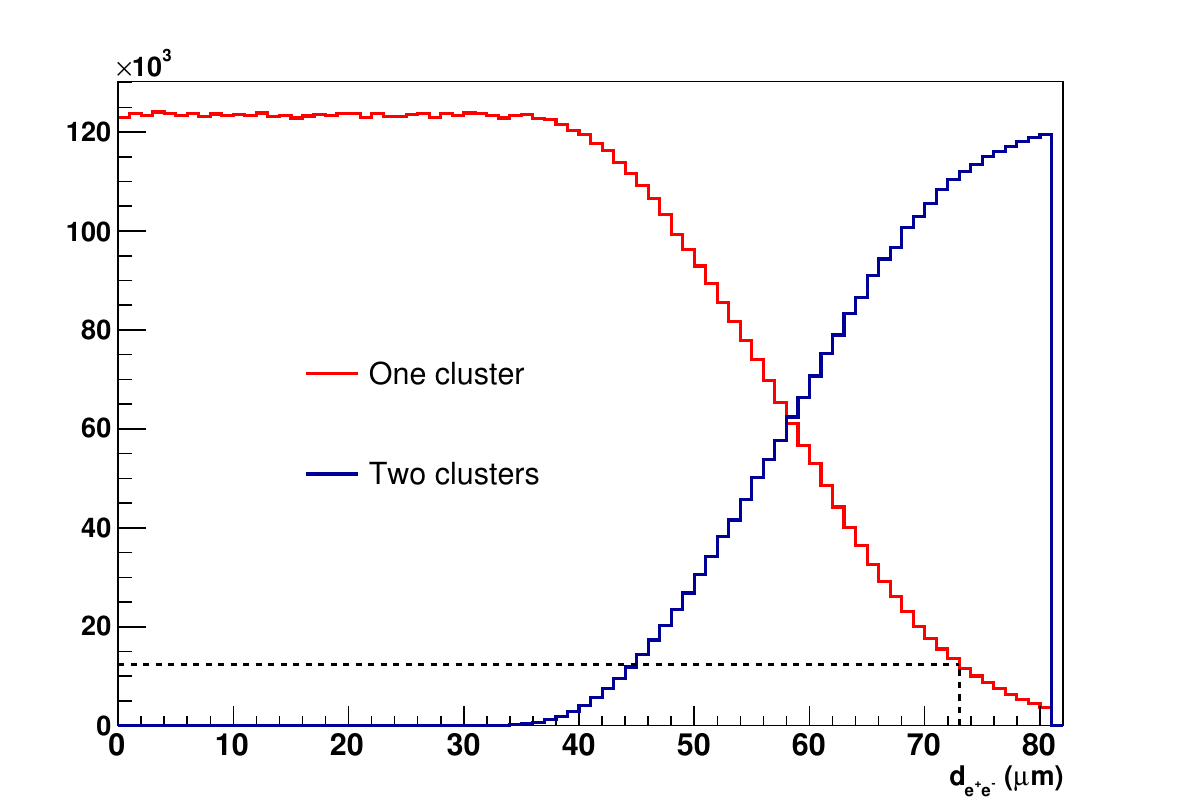}
  \caption{\label{fig:clus-dee} 1-cluster/2-cluster event configuration as function of the distance between the two particles in the simulation.}
\end{figure}

For this study, a pixel is considered part of a cluster if it shares a side with a pixel belonging to this cluster. Consequently a pixel A at (column N,row P) and a pixel B at (column N+1,row P+1) are considered two single-pixel clusters. These rules implemented in a simple cellular automaton for clustering, the number of detected clusters as function of the distance between the two particles is plotted in Figure~\ref{fig:clus-dee}. When the particle are separated by less than 36~$\mu$m, corresponding to approximately 1.5 pixel, a single cluster is always detected. Around 57~$\mu$m, it is a 50\%/50\% chance to get either one or two clusters. When the distance is greater than 73~$\mu$m apart, the likelihood for a single-cluster drops below 10\%.   

Although the ALPIDE sensor does not provide the collected charge in the pixel, the shape of the 1-cluster or very-close-2-cluster configuration leads to an accurate determination of the particle positions as shown in the next subsection.

\subsection{Cluster shape analysis and modeling}

\subsubsection{Shape analysis}
\begin{figure*}[!hbt]
  \centering
  \includegraphics[width=0.249\linewidth]{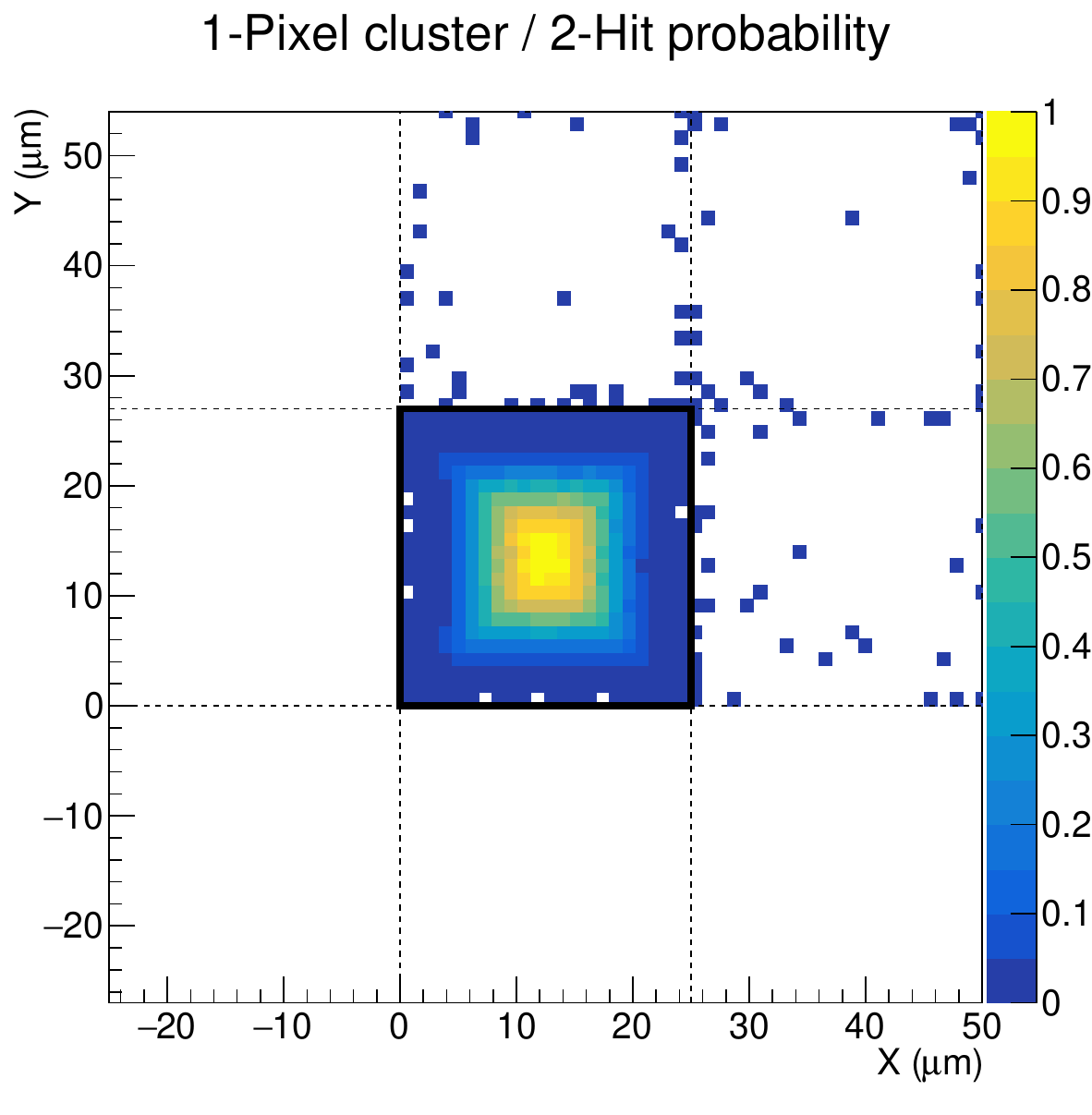}\\
  \includegraphics[width=0.249\linewidth]{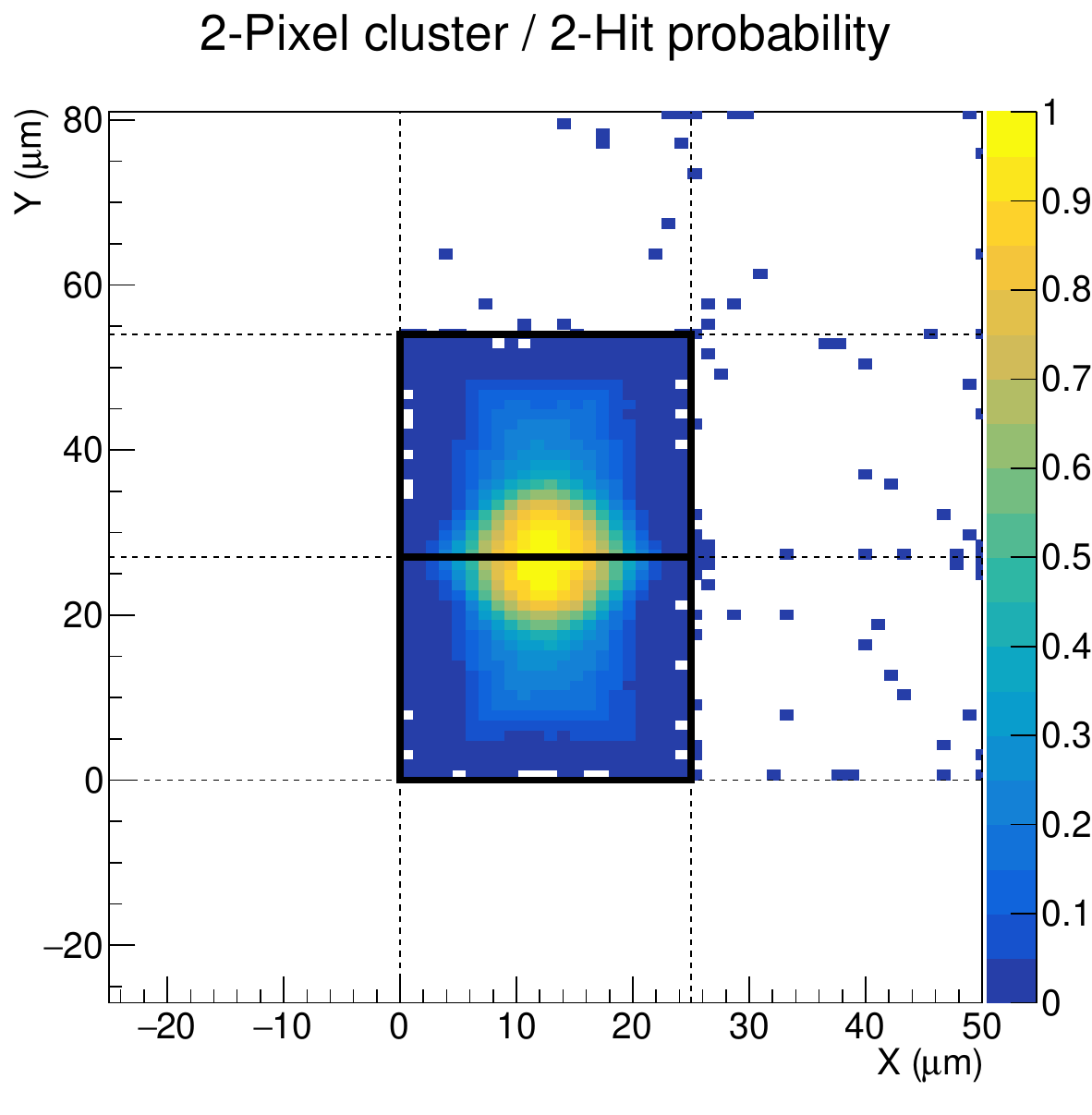}\includegraphics[width=0.249\linewidth]{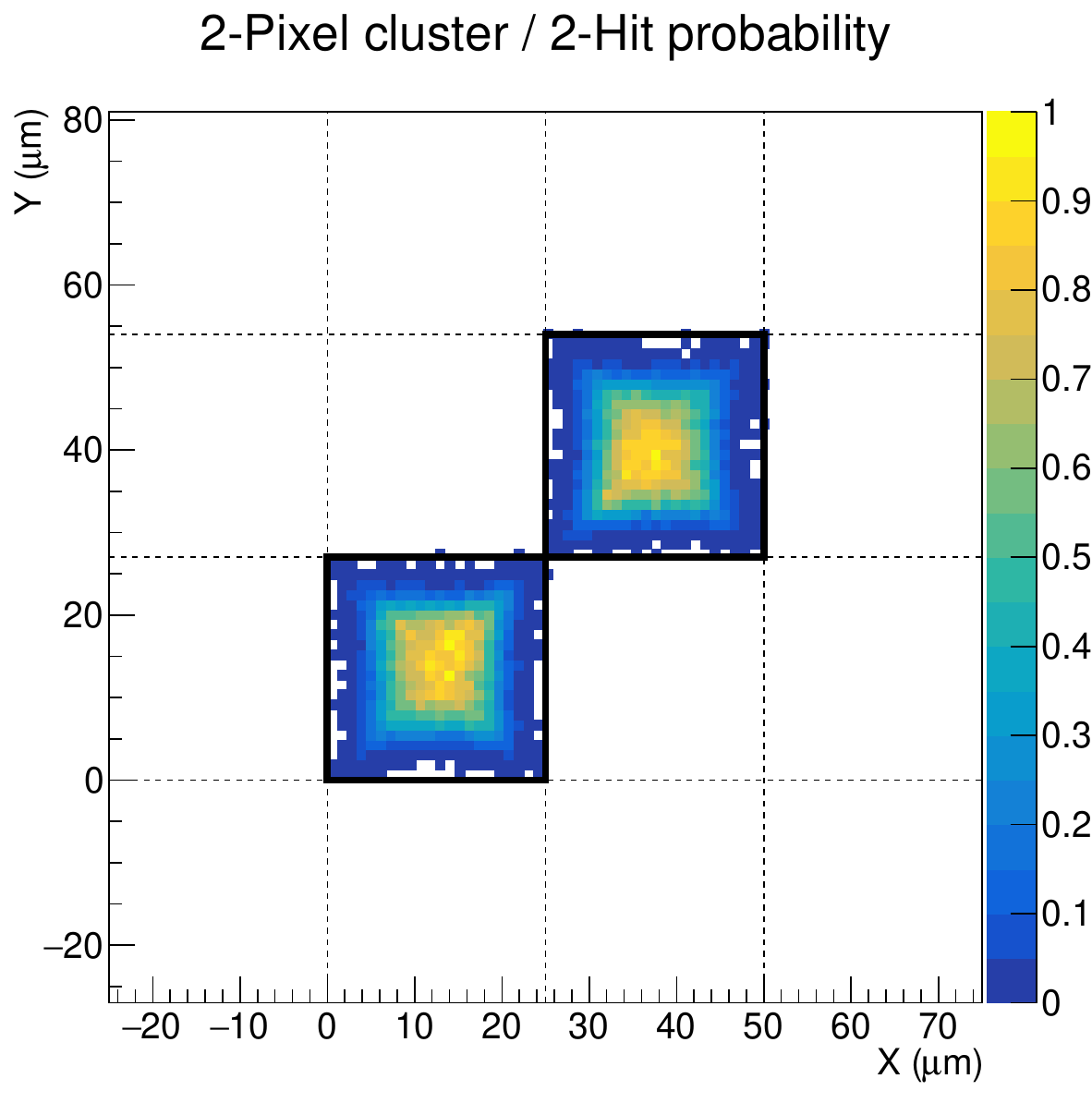}\\
  \includegraphics[width=0.249\linewidth]{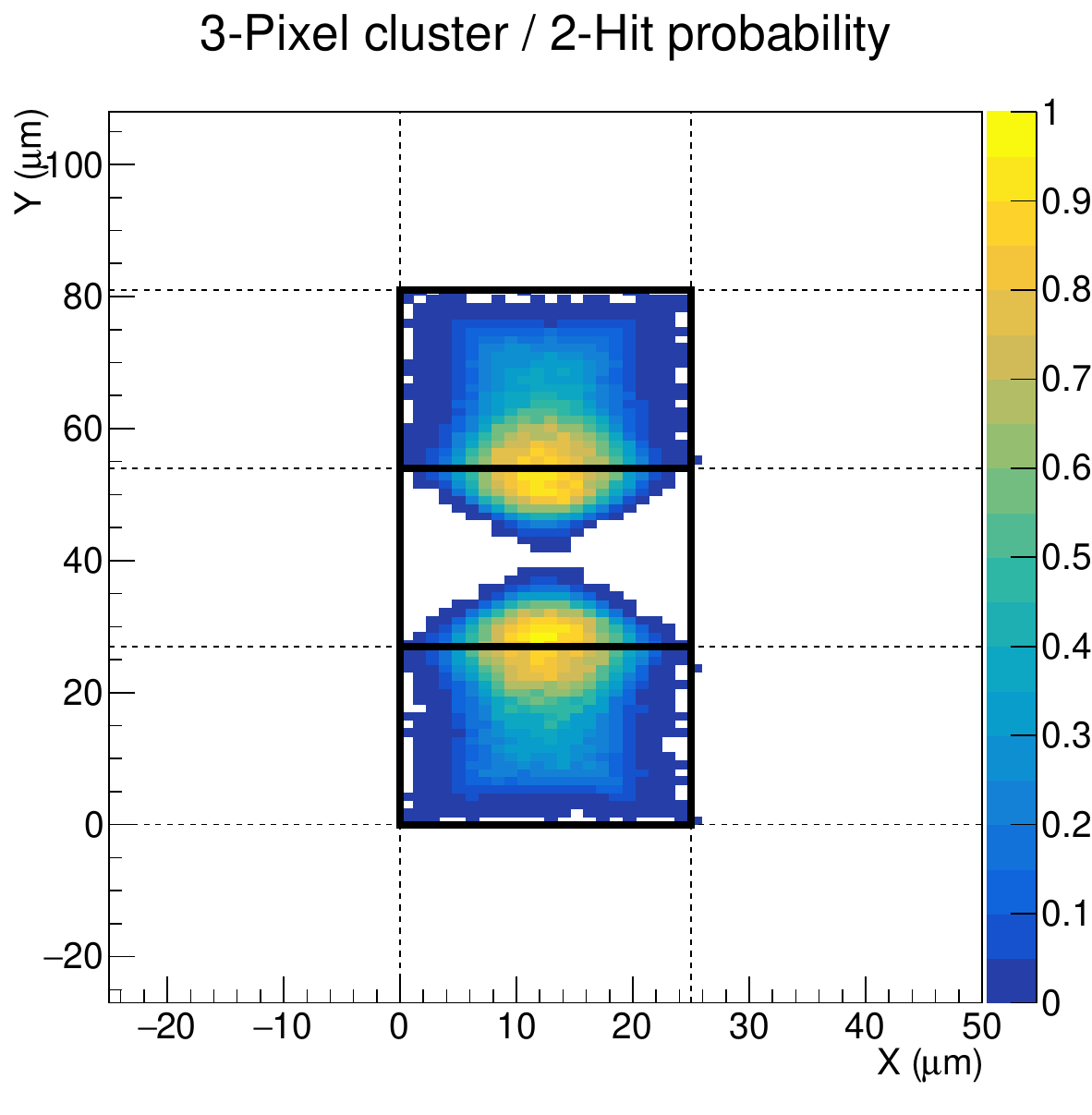}\includegraphics[width=0.249\linewidth]{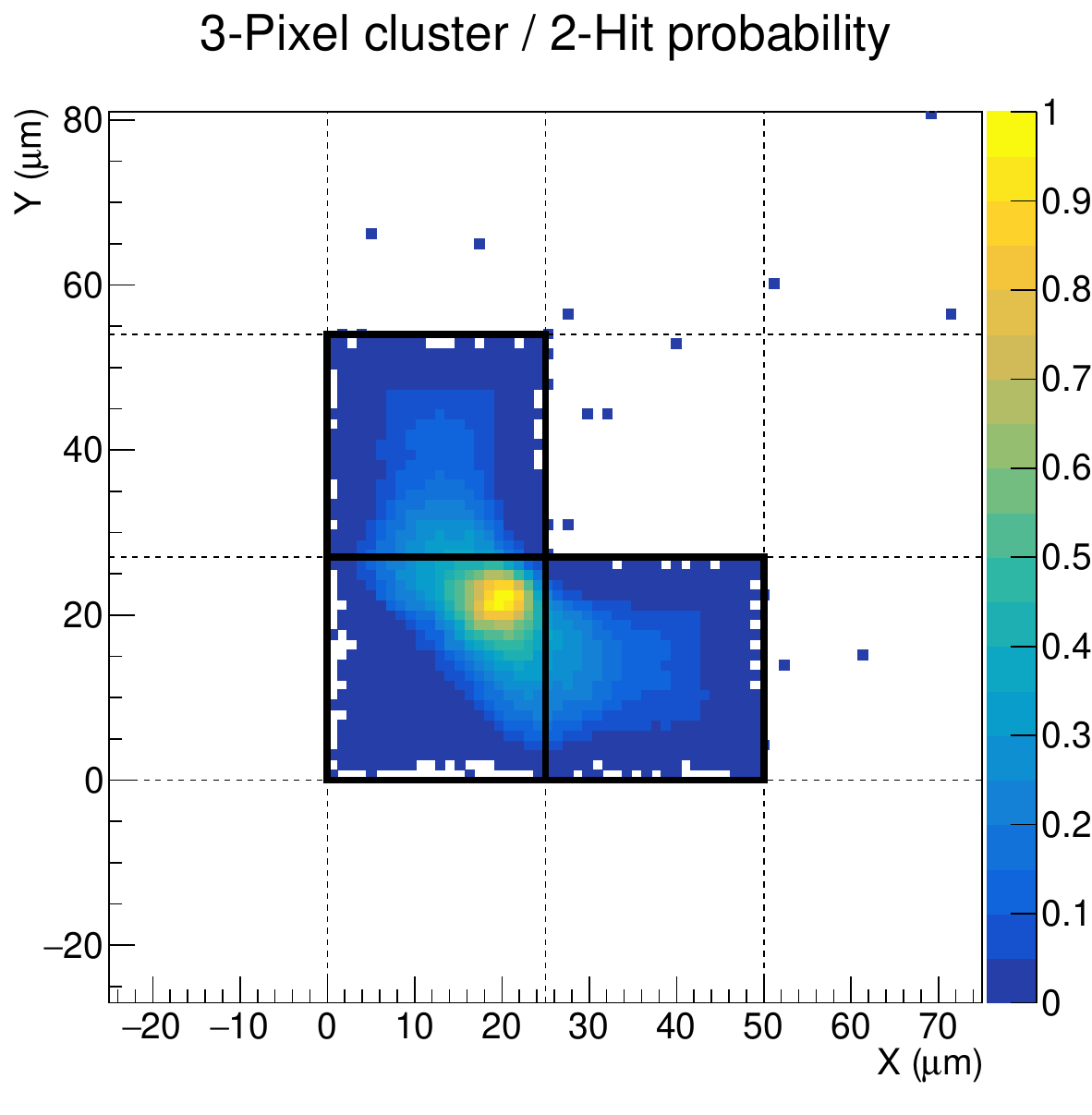}\includegraphics[width=0.249\linewidth]{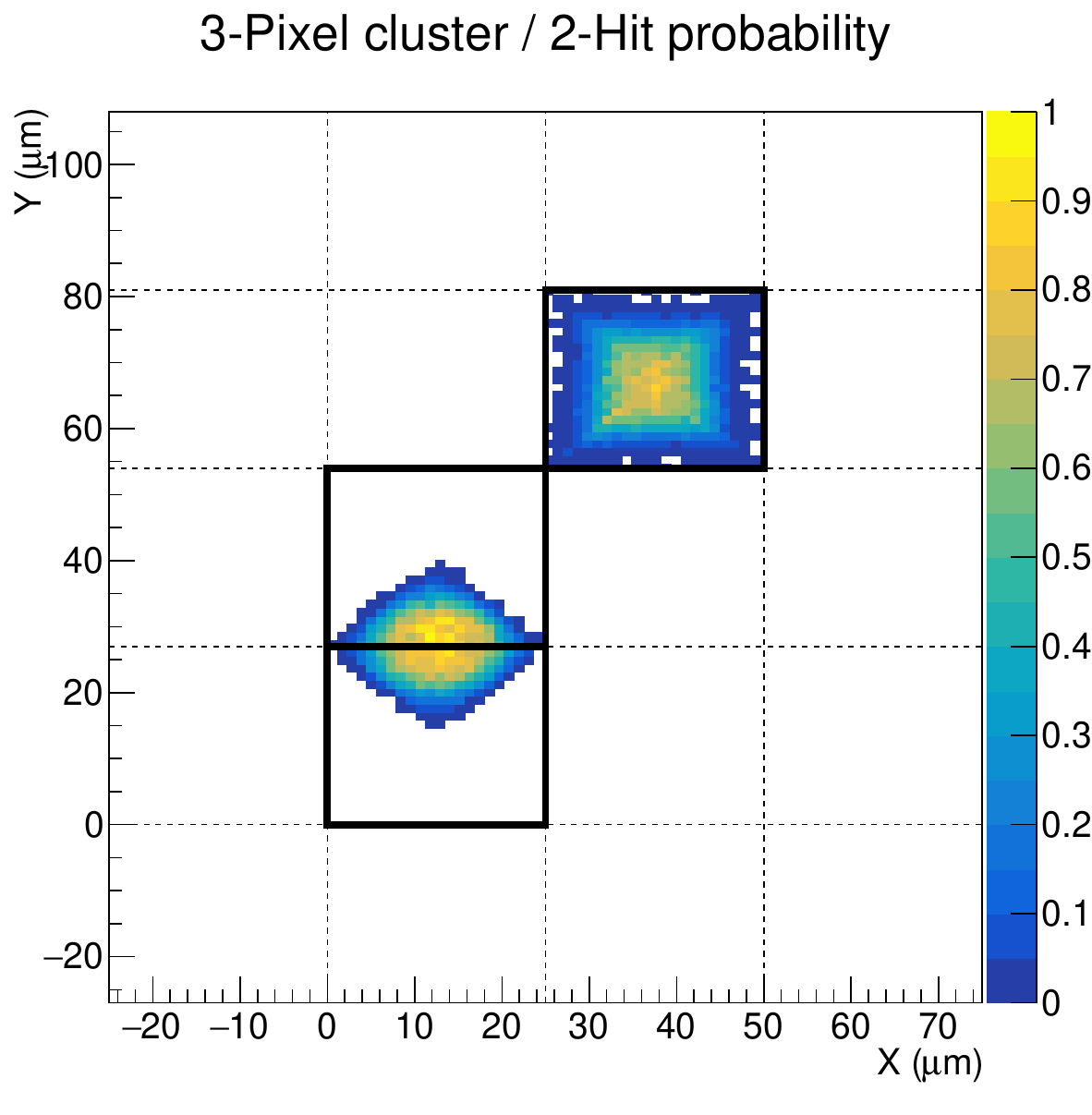}\\
  \includegraphics[width=0.249\linewidth]{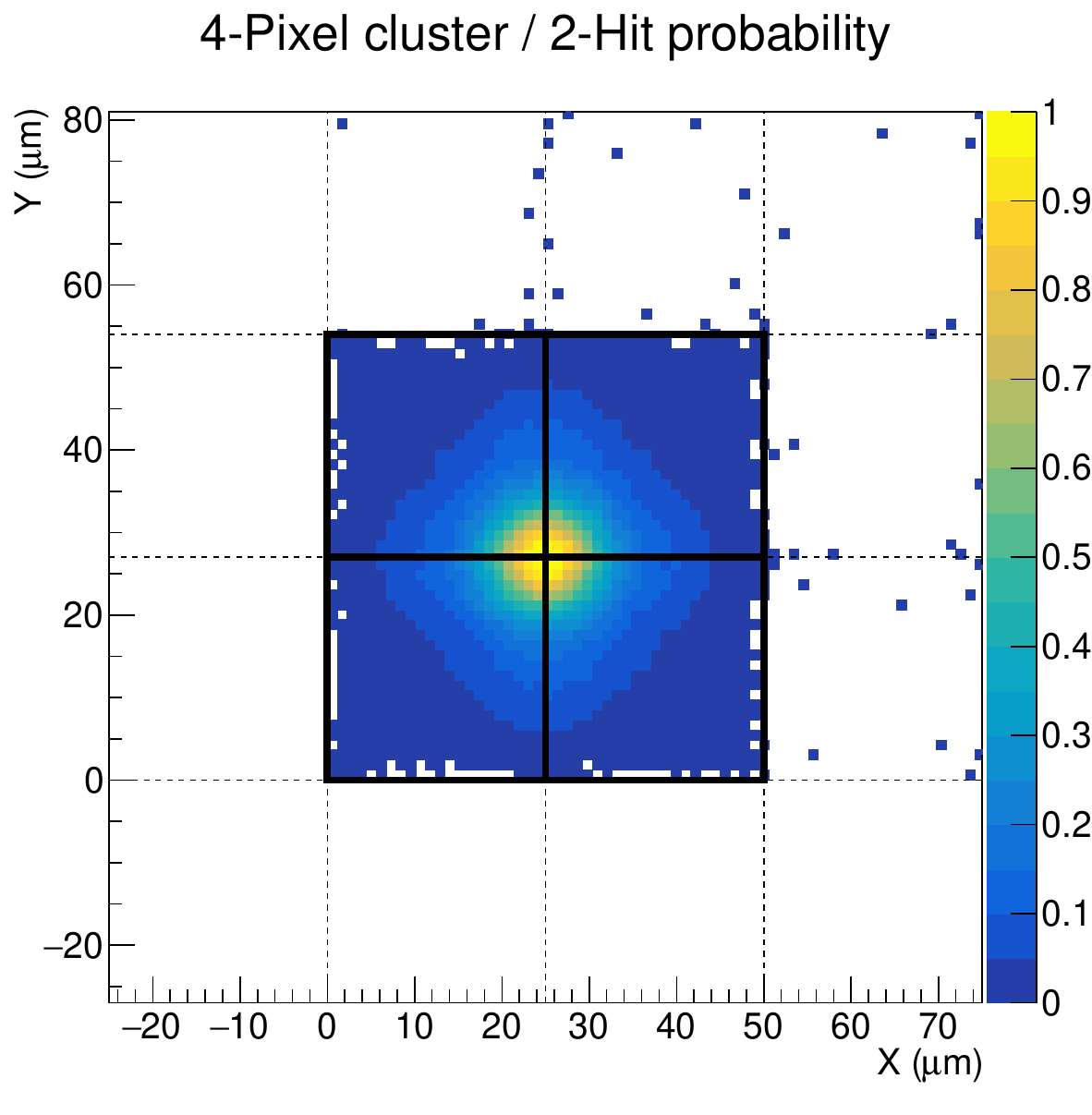}\includegraphics[width=0.249\linewidth]{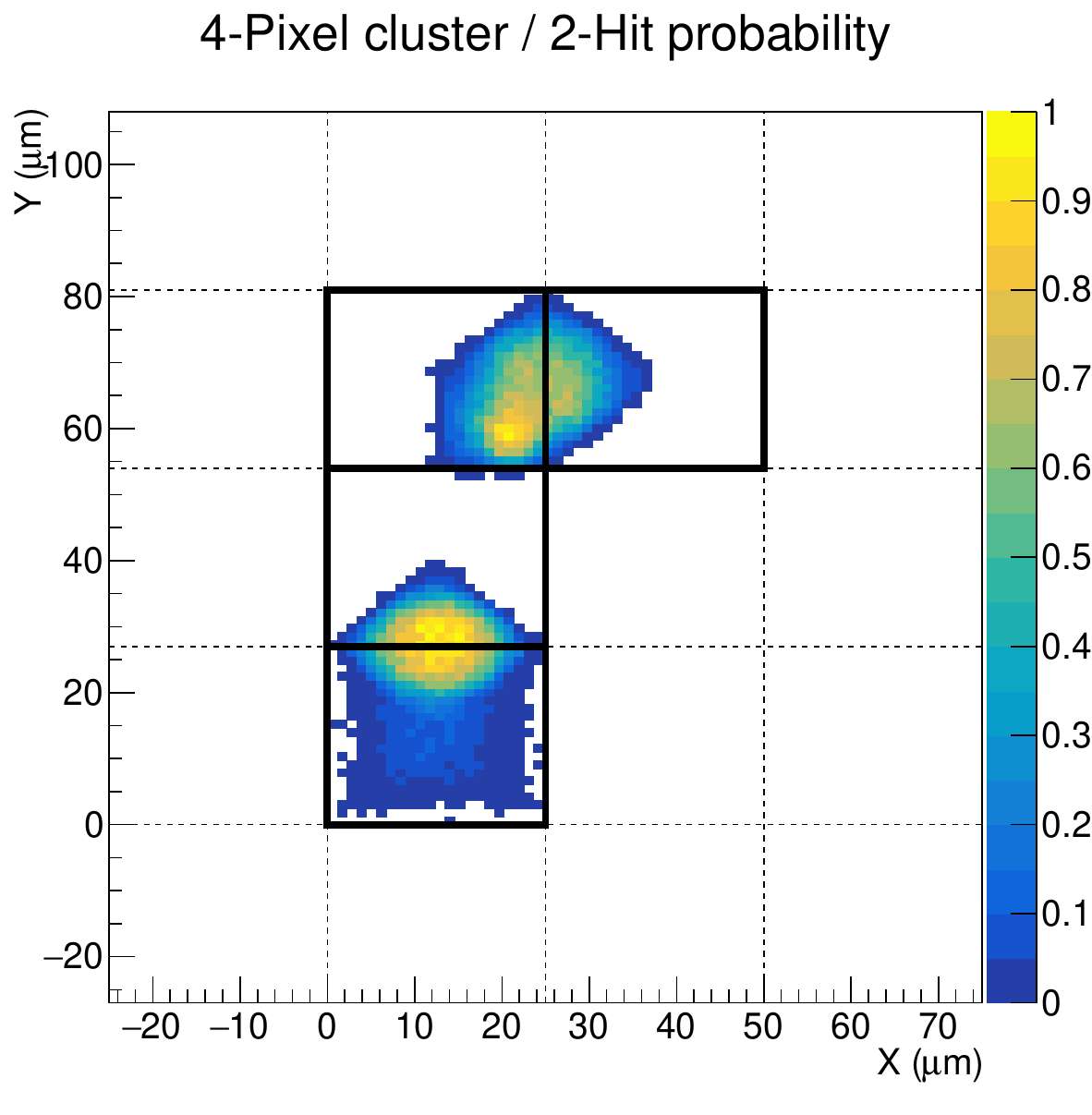}\includegraphics[width=0.249\linewidth]{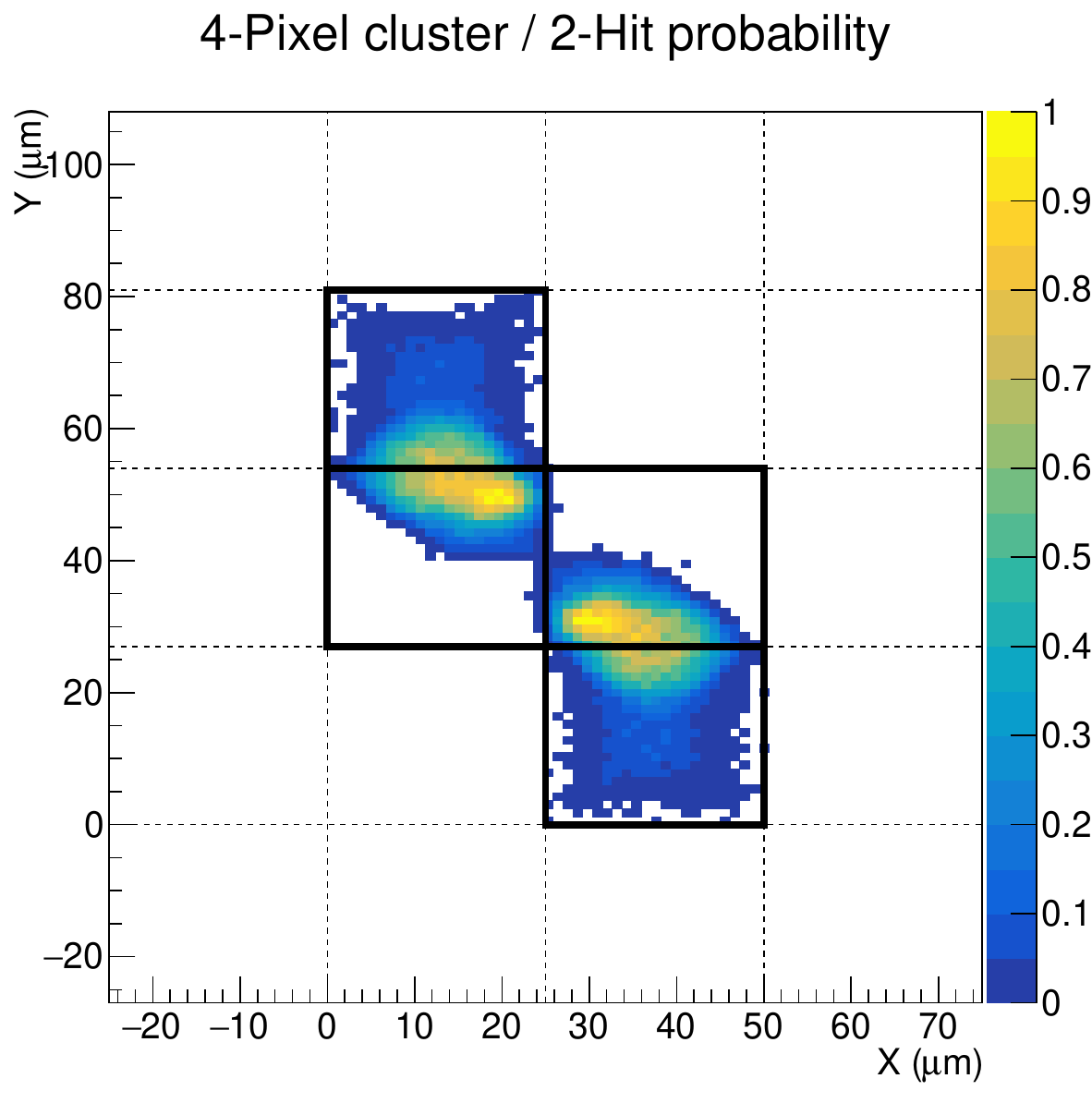}\includegraphics[width=0.249\linewidth]{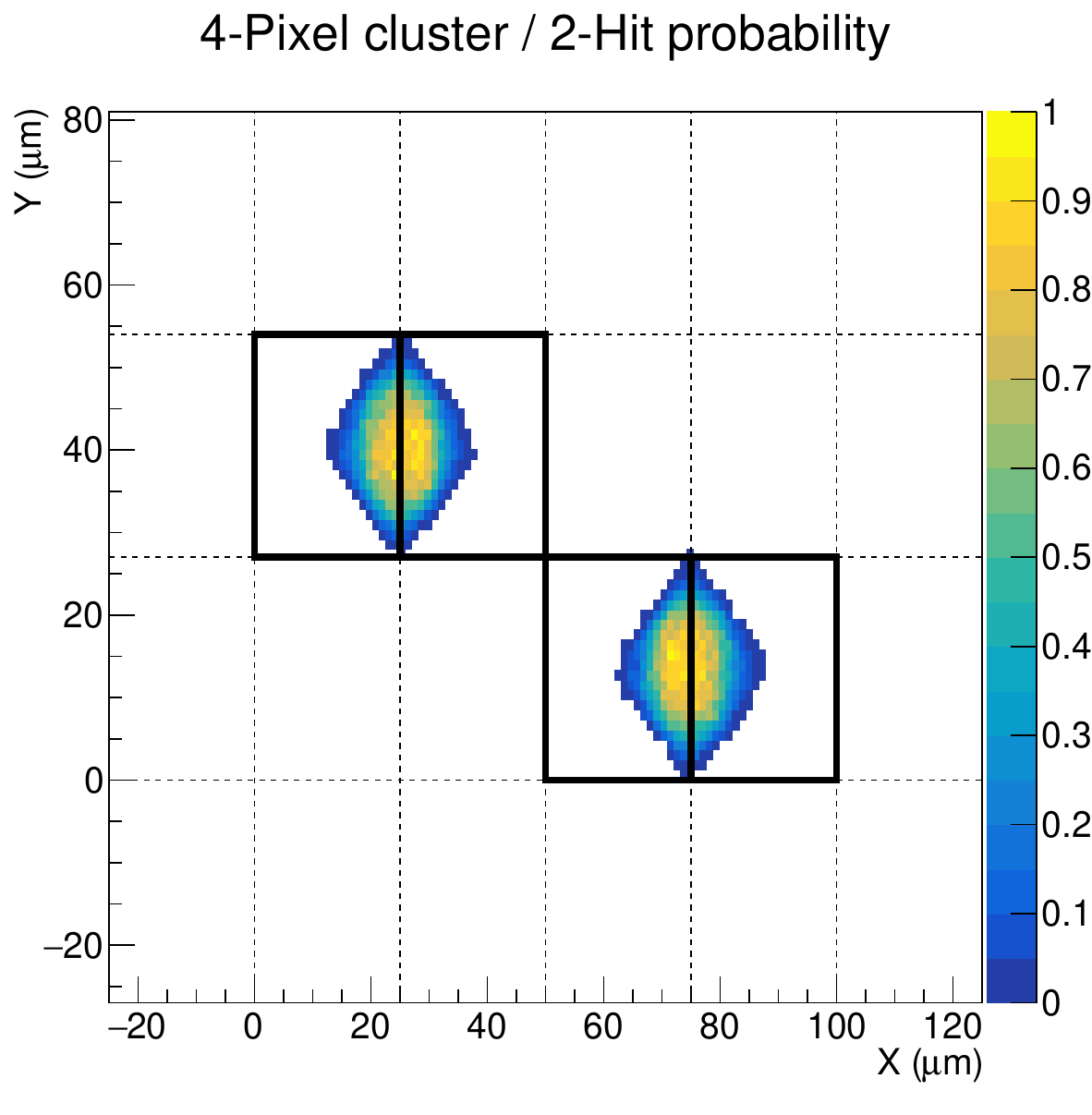}\\
  \caption{\label{fig:SomePix} Various cluster shapes and the associated X/Y 2-particle distributions.}
\end{figure*}

The cluster shape is not only given by the particle position but by their energy deposit as well. Using the Monte-Carlo simulation, it is possible to extract distributions of the 2-particle positions producing a given pattern in the MAPS layer.\\
The simplest shape is the single-pixel cluster pictured at the top of Figure~\ref{fig:SomePix}. Looking at the associated distribution of particle positions, the most likely configuration is given by the two particles hitting the middle of the pixel together. Indeed, if one particle hits the pixel close to an edge, the associated energy deposit cannot be too large as it may trigger the neighbor pixel. In this configuration, both particle positions can safely be assumed in the middle of the pixel with an excellent resolution. \\
The second row (from the top) of Figure~\ref{fig:SomePix} illustrates the various 2-pixel cluster shapes. When the two pixels are on the same column (or row respectively), the most likely position of both particles is in the middle of the common pixel edge. If both pixels just share a corner, then the particles have gone through the center of each pixel. Indeed, a particle getting too close from a pixel edge would fire a third pixel, leading to a shape presented in the third row of Figure~\ref{fig:SomePix}. \\
When all three pixels are on the same column (or row), two independent positions are clearly observed. The 3-pixel L-shape cluster mostly results from the two particles going through the common corner of all pixels. The last 3-pixel configuration is formed of two clusters providing an accurate position of each particle.\\
The bottom row of Figure~\ref{fig:SomePix} is dedicated to the 4-pixel configurations that perfectly illustrate the complexity of the underlying distributions for the particle positions. The configuration leading to the worst resolution of all possibilities is the $2\times 2$-pixel shape. The second, third and fourth 4-pixel configuration exhibit all 2 well-identified distributions for each particle with complex shapes. \\

\subsubsection{Modeling the position distributions}
\label{ssec:shape}
Knowing the underlying 2-particle distributions of a cluster shape is interesting but useless if not properly used in a Kalman filter. As the reader probably knows, assuming that noise and uncertainties are following gaussian distributions is convenient as it provides a standard deviation. This standard deviation is then injected in the Kalman Filter to compare the reliability of the prediction against the reliability of the measurement in order to derive the best estimate of the particle position. But many of the position distributions cannot be considered gaussian.\\

\begin{figure*}[!hbt]
  \centering
  \includegraphics[width=\linewidth]{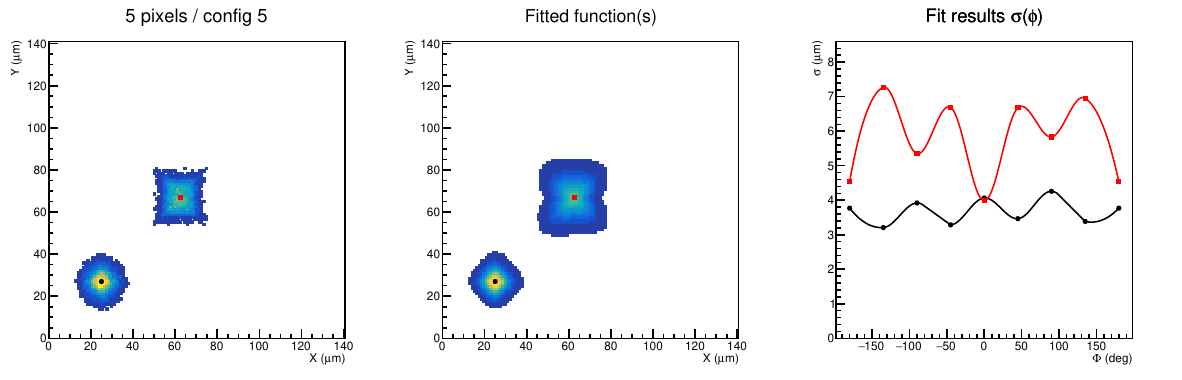}\\
  \includegraphics[width=\linewidth]{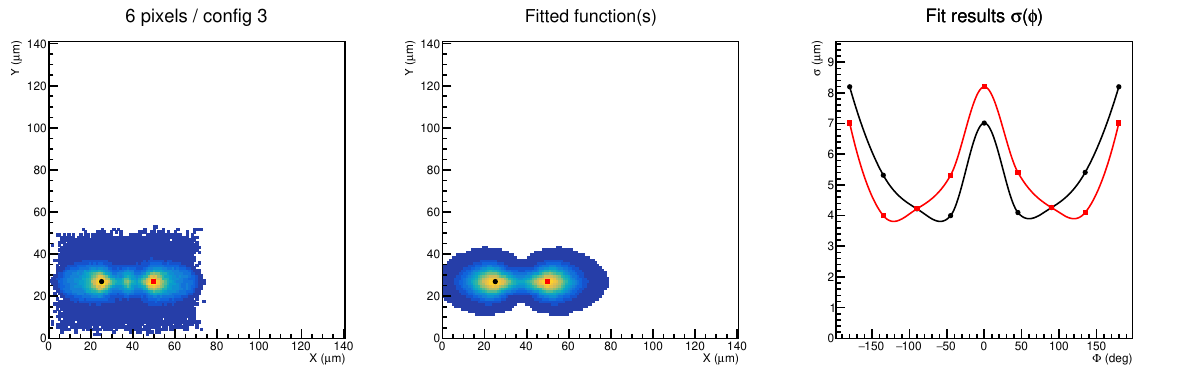}\\
  \caption{\label{fig:fit_shapes} Examples of cluster shapes and associated fits. The left column represents the original distributions to be fitted. The middle column represents the fitted distributions with black and red markers indicating their centers. The right column displays $\sigma(\phi)$ for one or two distributions.}
\end{figure*}

Consequently it was decided to use the following function to fit the position distribution:
\begin{eqnarray}
  f(x,y)&=&\mathbf{C} \times \exp\left(-\frac{\sqrt{(x-\mathbf{x_{_0}})^2+(y-\mathbf{y_{_0}})^2}}{2 \mathbold\sigma^2(\phi)}\right)\;,\\
  \phi&=&\textrm{atan}(y-\mathbf{y_{_0}},x-\mathbf{x_{_0}})\;
\end{eqnarray}
where $\mathbf{C,x_{_0},y_{_0}}$ are three parameters to be fitted. Regarding $\mathbold\sigma(\phi)$, it represents an array of parameters to model the skewness of the distribution around the distribution maximum. Figure~\ref{fig:fit_shapes} illustrates the resulting fits for various distributions.

Approximately 97\% of all possible configurations are fitted, with cluster size from 1 to 8 pixels, and saved in a file to be later used in the reconstruction. The remaining 3\% are mostly composed of all possible configurations from 9 to 14 pixels, possible only by exceptionally large amount of energy deposited by the two particles in the silicon. If such configuration is generated in the Monte-Carlo simulation, it would be discarded and regenerated until the cluster matches a configuration saved in the file.

\section{Reconstruction algorithm for a multi-GeV photon polarimeter}
A full simulation-reconstruction code has been written so that all experimental parameters could be tested. The list of parameters includes the cell geometry, the number of cells, the density of the Rohacell-like converter, the photon energy and its polarization. Conversions are generated in the first layer of converter and Geant4~\cite{GEANT4} takes over the propagation of the electron/positron pair in the various materials. When crossing the MAPS, the position of both particles in the middle of the silicon plane are stored in a tree. Secondary particles stemming from the interaction of the lepton pair with the matter are ignored in the present study.\\

If the distance between two particles is found larger than 73~$\mu$m, their positions are independently smeared by 4~$\mu$m. However the full simulation of the energy deposit in the MAPS layer for both particles is performed when they are closer than 73~$\mu$m. The resulting cluster pattern is then identified and the associated position distributions are saved as measurements to be used later in the Kalman filter. 

\subsection{Reconstructing the two tracks and the azimuthal angle of the conversion}
\label{ssection:azirec}
A Kalman filter is run backwards, \emph{i.e.} from the last layer to the first layer encountered by the lepton pair. Two track candidates are initiated, one for each measurement in the last layer. The position of the particle is set by the measurement and the tangent of the track angle $tan\left(\theta\right)$ is initialized with the photon angle (assuming a PbWO$_4$ calorimeter is installed behind the pair polarimeter). The position uncertainty is set to the measurement accuracy (4~$\mu$m) and a multiple of the mean deviation by Coulomb scattering sets the uncertainty on the track angle.\\

Then the two state vectors are propagated to the next layer. The distances between the projected positions and each measurements are computed. The most likely combination of tracks and measurements is chosen by choosing the combination minimizing the sum of the squared distances. In other words, the most likely combination is picked. Filtering is performed and propagation to the next layer is then run. In case of a measurement being a distribution, the measurement uncertainty is estimated with the angular position of the projected position with respect to the center of the distribution provided by $\mathbold \sigma\left(\phi\right)$ introduced in subsection~\ref{ssec:shape}. \\

Once the filtering on the very first measurement(s) is performed, the distance between the center of the two distributions is used to assess the depth of the conversion in the converter. We know that the two tracks must cross at a given depth in the converter. Consequently the two track candidates are propagated to this depth and the projected position of one lepton is used as an extra measurement to filter the state vector of the other lepton. The uncertainty assigned to this last measurement is partially given by the propagated covariance matrix combined with the uncertainty on our guess of the conversion depth. \\

Once this final filtering step is performed in the converter, the state vector needs to be smoothed at the first MAPS layer to get the best estimate of both particle positions. From state vectors at the converter and at the first MAPS layer, there are two ways to compute the azimuthal angle of the conversion:
\begin{enumerate}
\item Using the track angles at the conversion location later referred as \emph{direction-based},
\item Or computing the track directions with the particle positions at the assumed conversion depth and at the first MAPS layer, called \emph{position-based} in the following sections.
\end{enumerate}

Both quantities have different uncertainties derived using the covariance matrices of the various state vectors. The final azimuthal angle is chosen to be the value associated to the smallest uncertainty.\\

\subsection{Extracting the effective analyzing power}

\begin{figure}[!hbt]
  \includegraphics[width=0.495\linewidth]{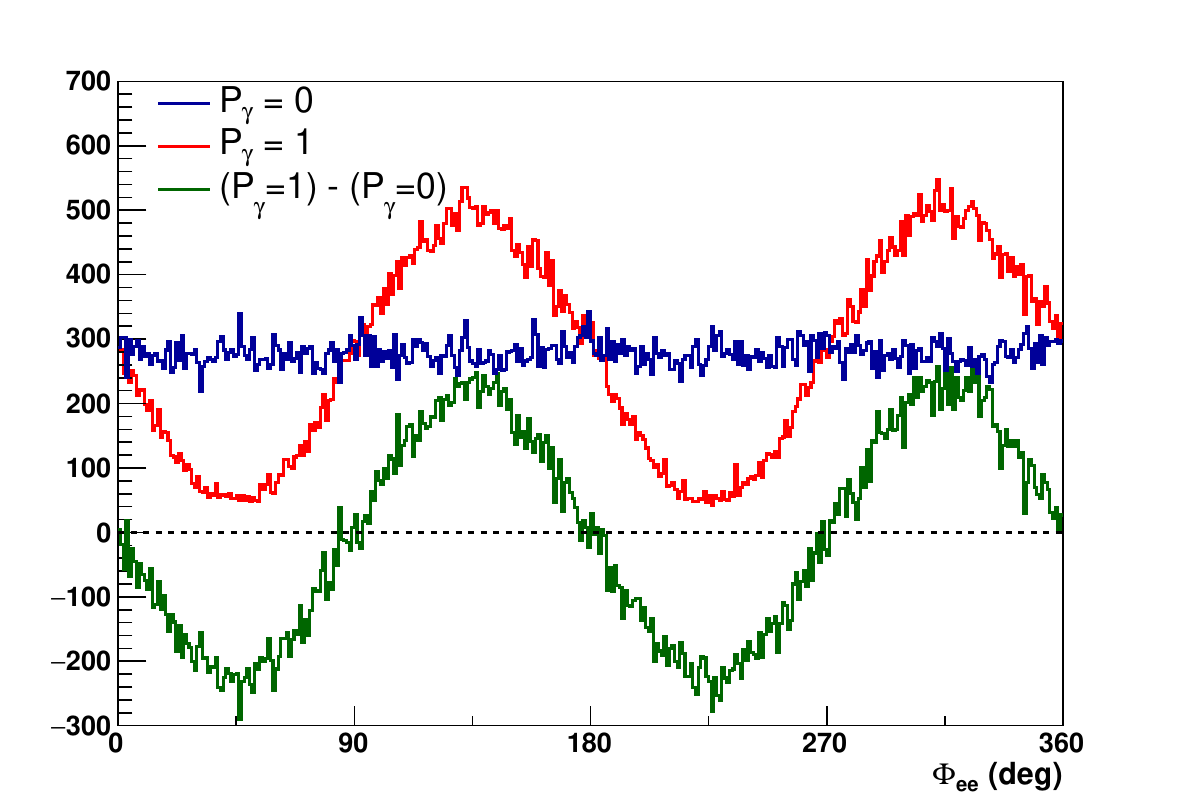}
  \includegraphics[width=0.495\linewidth]{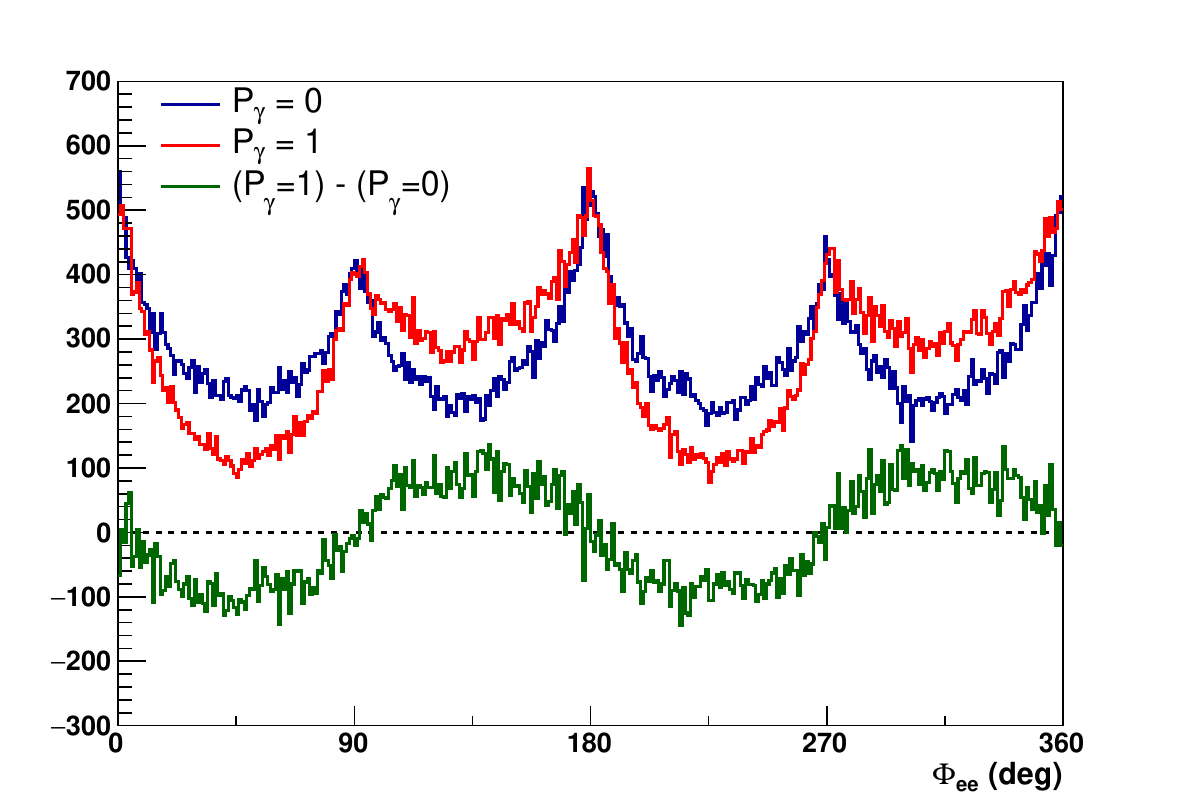}
   \caption{\label{fig:Pol0sub} Azimuthal angle obtained for 2-GeV photons with a 10cm-long cell length (left) and for 6-GeV photons with a 1cm-long cell. Blue distributions are obtained when the photon is not polarized, red ones with 100\%-linearly polarized photons. The analyzing power is obtained by fitting the amplitude of the green distribution resulting from the subtraction of polarized distribution with the unpolarized one.}
\end{figure}

Extracting the intrinsic analyzing power of a polarimeter design may be more complicated than just fitting a $\cos(2\phi)$-modulation in the reconstructed azimuthal distributions of the conversions. In Figure~\ref{fig:Pol0sub}, such distributions are shown for two different cell layouts and two different photon energy. The left plot displays reconstructed distributions for unpolarized (blue) and 100\%-linearly polarized (red) 2-GeV photons with MAPS spaced by 10~cm. For this configuration, the MAPS resolution is negligible compared to the multiple scattering and the distance between the two leptons. Consequently the intrinsic analyzing power can be straightfowardly extracted by fitting the distribution with $B \times \left(1+A_I\cos(2\phi)\right)$.
The right plot in Figure~\ref{fig:Pol0sub} shows the same distributions except that the distance between two consecutive MAPS planes is 1~cm with 6-GeV photons. For this configuration the pixel size is now comparable to the distance between the two leptons or displacements induced by multiple scattering. As a consequence the distributions exhibit spikes at 0/360-,90-,180- and 270-degrees corresponding to \emph{undetermined} measurements of the angle due to the pixel size. Comparing the position of the polarized distribution compared to the unpolarized one, the $\cos(2\phi)$-modulation is noticeable and fully revealed by the green distributions resulting from the subtraction of the unpolarized distribution off the polarized one.\\       
 
Therefore, to properly extract $A_I$ in the next sections, two simulations are run with unpolarized and polarized photons. Then the reconstruction is run on both simulations and $A_I$ is fitted on the difference between unpolarized and polarized distributions. 

\section{Optimization of the polarimeter design} \label{sec:optimization}
Optimization of the polarimeter is performed for photon energies of 2, 4 and 6~GeV. First the dependence of the intrinsic analyzing power with respect to the cell length is assessed for a thin converter. Then, still with a thin converter, the dependence of the analyzing power with respect to the position of the MAPS within a cell is tested. Finally, in the last section, density and length of the converter are fine-tuned to maximize the Figure-of-Merit of the polarimeter. For all next subsections, the polarimeter is composed of 16 cells, all used in the Kalman filter to reconstruct the conversion azimuthal angle. The MAPS planes are all 50$\mu$m-thick for the next plots. 

\subsection{Cell length}
\begin{figure}[!hbt]
   \includegraphics[width=0.495\linewidth]{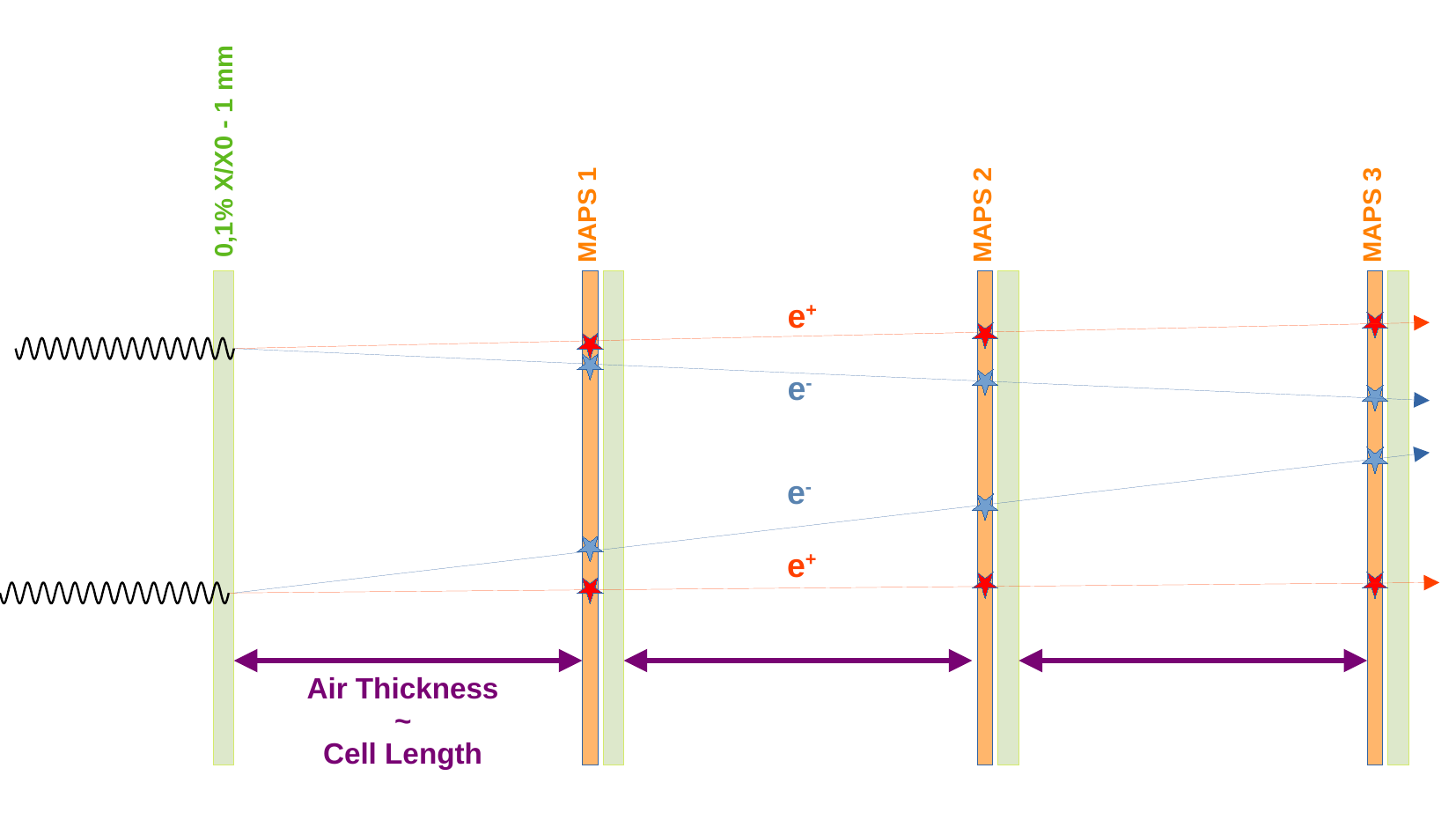}
   \includegraphics[width=0.495\linewidth]{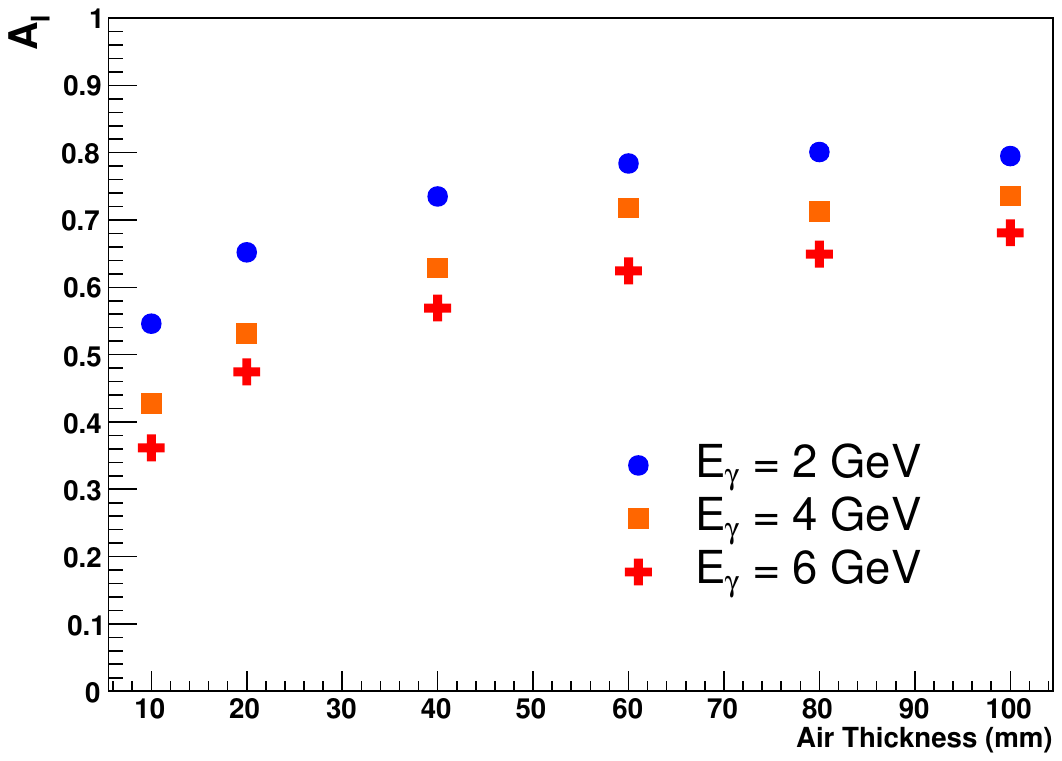}
   \caption{\label{fig:CellOpt} Left: Schematics of the cell design for the length optimisation. Right: Extracted A$_{_{I}}$ for 2-(blue), 4-(orange) and 6-(red) GeV photons for various cell lengths.}
\end{figure}

Here the cell is simply composed of a 1mm-converter of Rohacell-like material with a density of 40~kg/m$^3$, separated from a MAPS plane by a layer of air as described by Figure~\ref{fig:CellOpt}. Then the intrinsic analyzing power is extracted for various lengths of the air layer and shown in the right plot of Figure~\ref{fig:CellOpt}.

As expected, A$_I$ increases as the cell length increases. Indeed, all MAPS layers having the same resolution, the distance improves the constraint on the direction of both leptons. From this study, we conclude that a cell should not be shorter than 80~mm to study photons up to 6~GeV.

\subsection{Converter-MAPS distance}
For this next study, the same thin converter is used and the cell length is fixed at 100~mm. However the distance between the MAPS plane and the converter in the cell is changed from 1 to 80~mm.

\begin{figure}[!hbt]
\centering
  \includegraphics[width=0.495\linewidth]{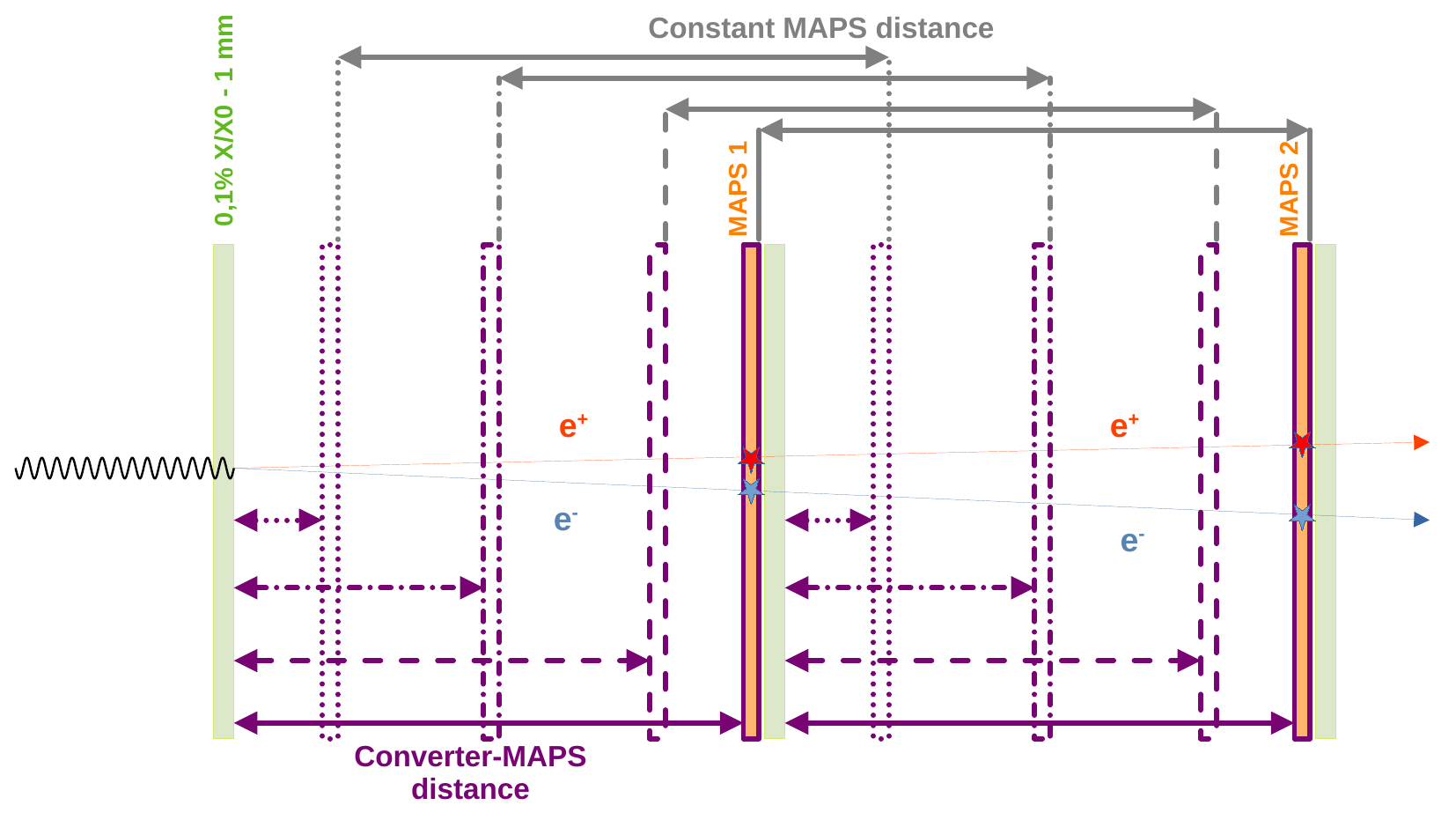}
  \includegraphics[width=0.495\linewidth]{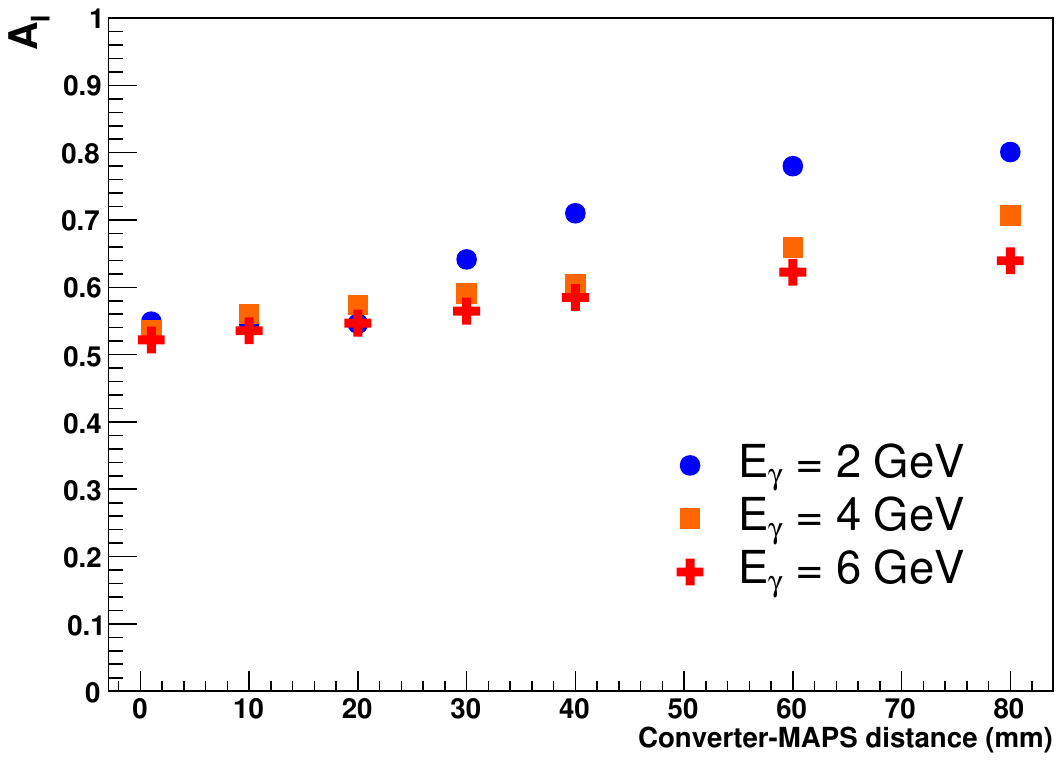}
  \caption{\label{fig:ConvMAPSPOS} Left: The distance between the converter and the next MAPS is changed whereas the distance between MAPS (and consequently converters) is kept constant at 100~mm. Right: Extracted A$_{_{I}}$ for 2-(blue), 4-(orange) and 6-(red) GeV photons for various Converter-MAPS distances.}
\end{figure}

\begin{figure*}[!hbt]
  \includegraphics[width=0.329\linewidth]{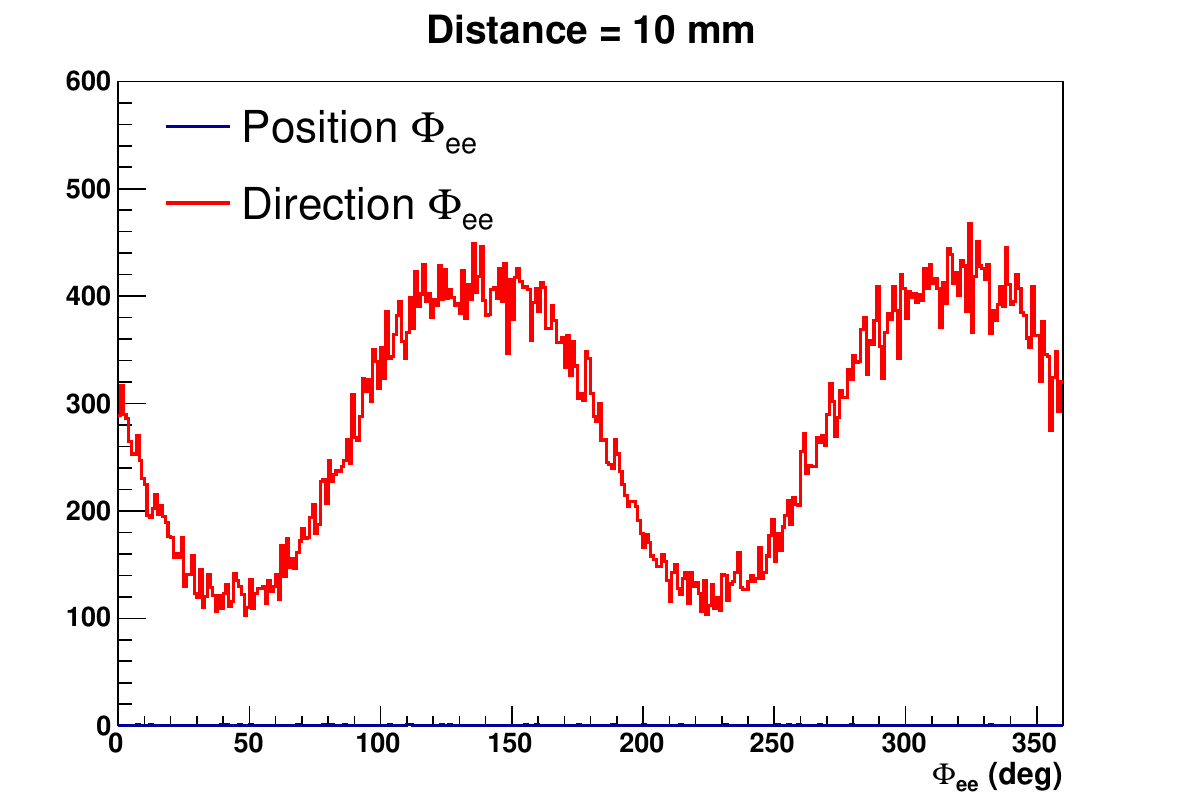}\includegraphics[width=0.329\linewidth]{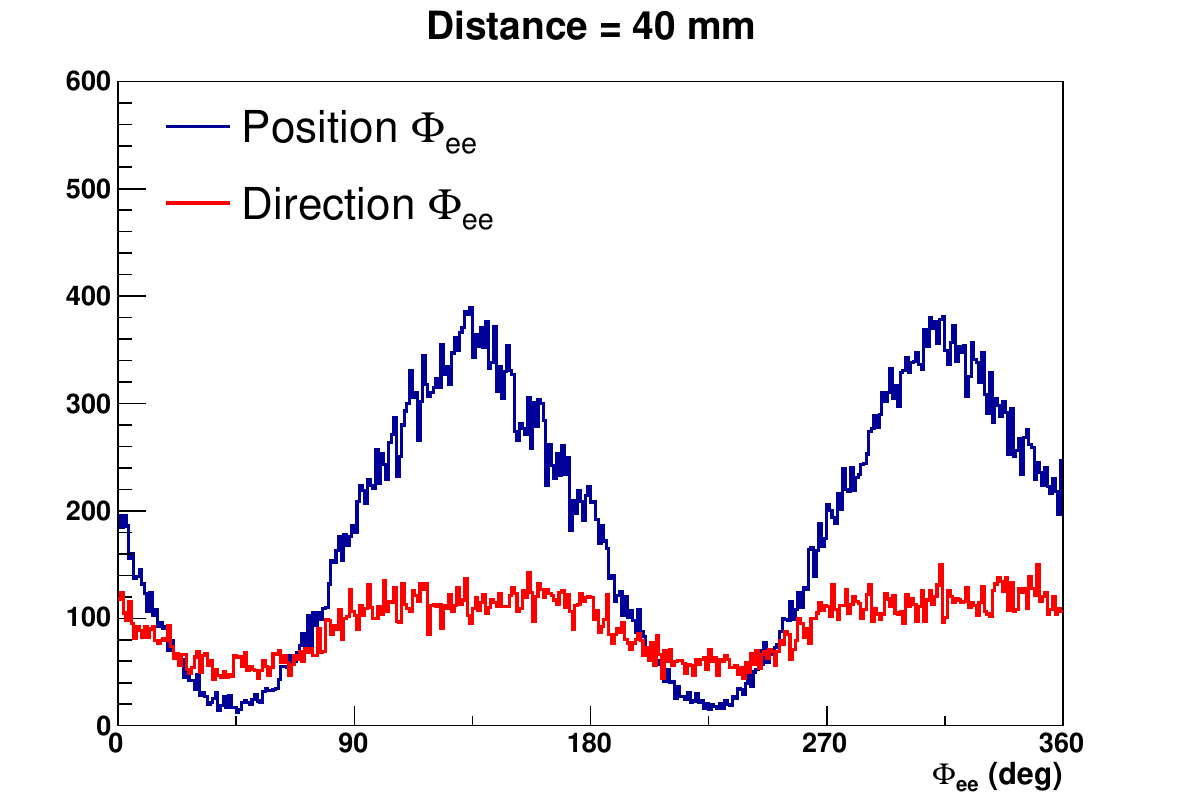}\includegraphics[width=0.329\linewidth]{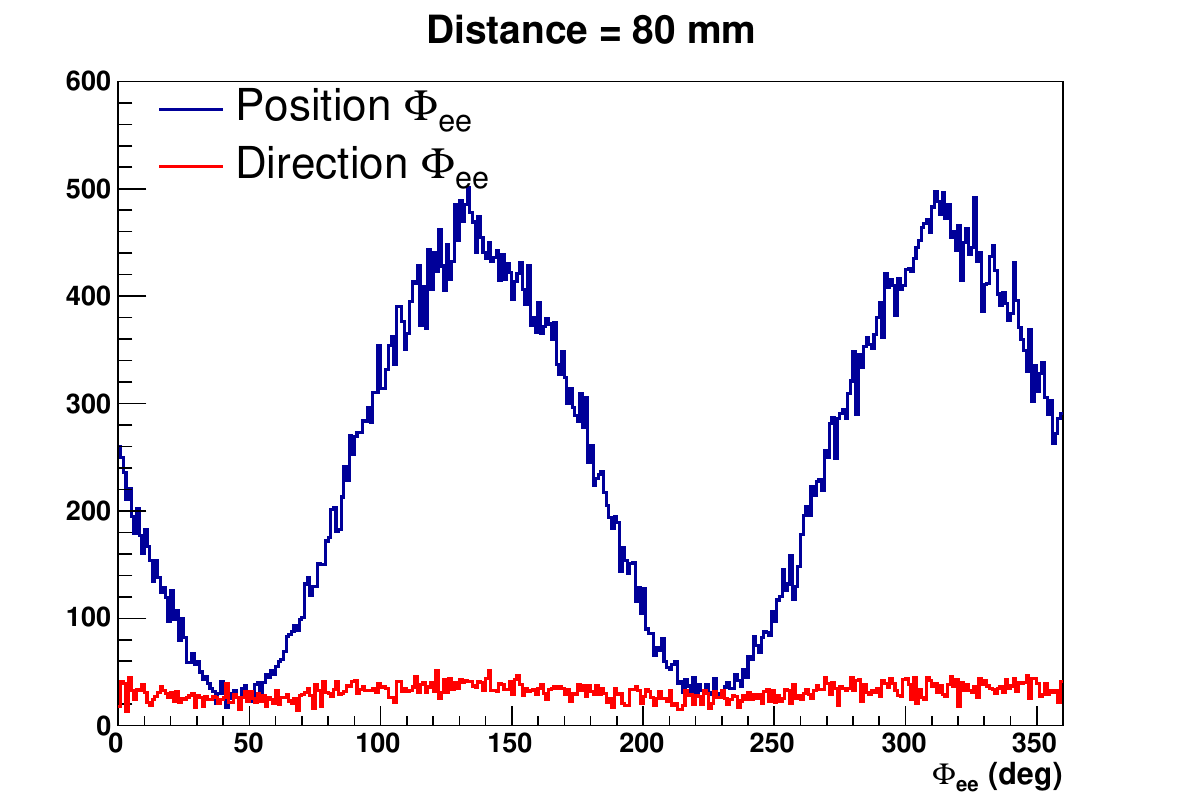}
  \caption{\label{fig:DirPos} Direction-based (red) and position-based (blue) azimuthal angle for converter-MAPS distance of 10 (left), 40 (middle) and 80~mm (right).}
\end{figure*}

A$_I$ appears to slowly increase with the Converter-MAPS distance for all three energies, although the distance between MAPS is constant. To understand this conclusion, a closer look must be taken on the computation of the azimuthal angle between the position-based and the direction-based methods introduced in subsection~\ref{ssection:azirec}. In Figure~\ref{fig:DirPos}, distributions of the direction-based (red) and position-based (blue) reconstructed azimuthal angles are shown as a function of the distance between the converter and the MAPS. These distributions are filled on an event-by-event basis, only when the associated uncertainty is found smaller than its alternative. 

For small distance between the converter and the next MAPS, the direction-based angle is the preferred quantity as it is mainly inferred from the two accurate measurements between the first two MAPS planes after the conversion. So close to the converter, the two leptons fire the same pixel in the first MAPS layer and the uncertainties on the extrapolated positions in the converter located a few millimeters behind are not small enough to gain any information.\\
On the contrary, the position-based quantity rapidly becomes the best estimate when the distance between the MAPS and the converter increases. This effect is particularly dramatic in this study as the converter is only 1-mm thick. Eventhough the uncertainty on the extrapolated position of one lepton in the converter is larger than a MAPS measurement, it is located far enough to provide a useful information to constrain the state vector of the second lepton (and vice versa). Moreover this extra-information does not come at the expense of more multiple scattering in a MAPS layer.\\

The conclusion of this study is that it is best to place the MAPS at the end of the cell. 

\subsection{Optimization of the radiation and converter lengths}
\begin{figure}[hbt]
  \centering
  \includegraphics[width=0.545\linewidth]{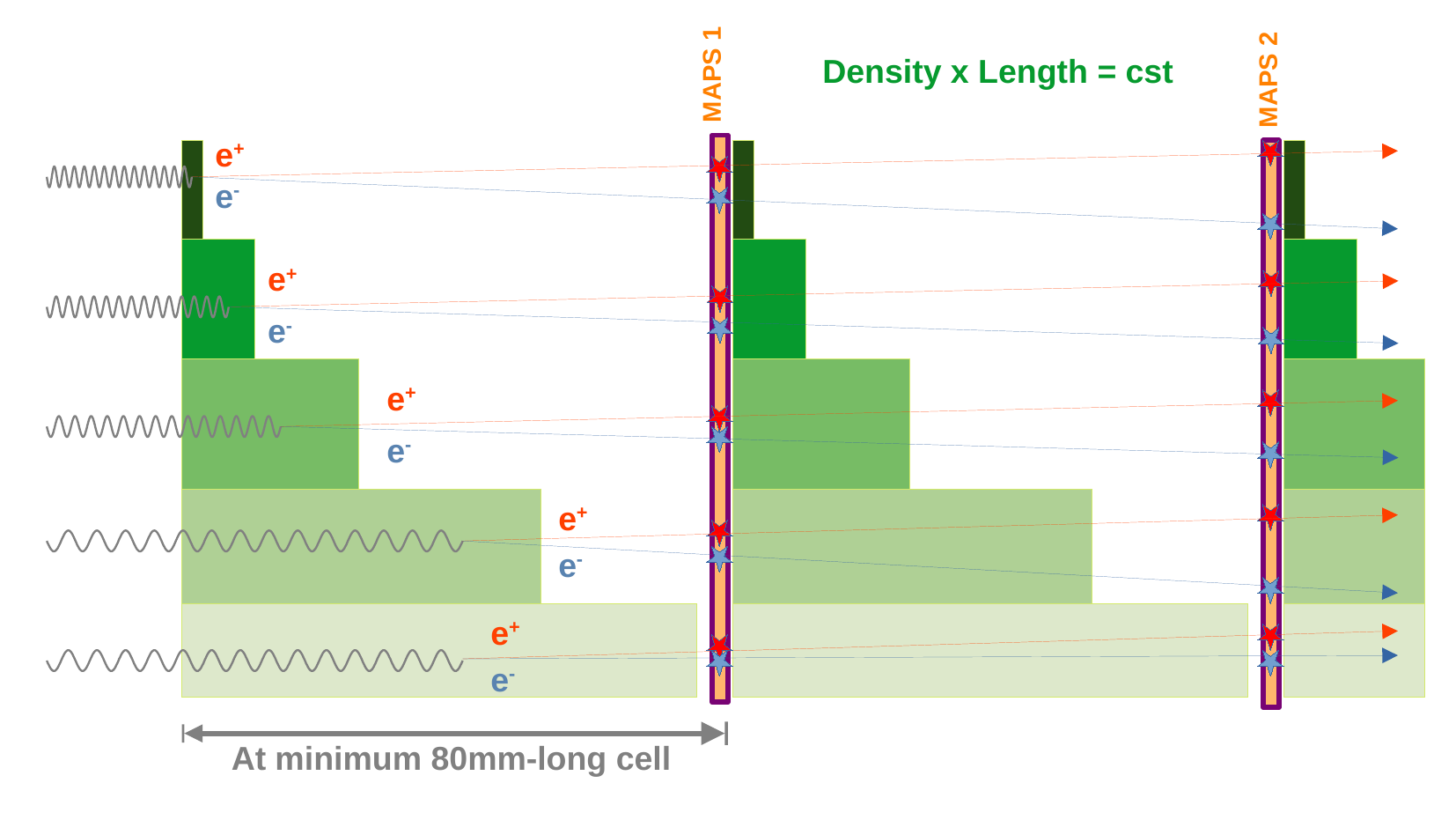} \includegraphics[width=0.445\linewidth]{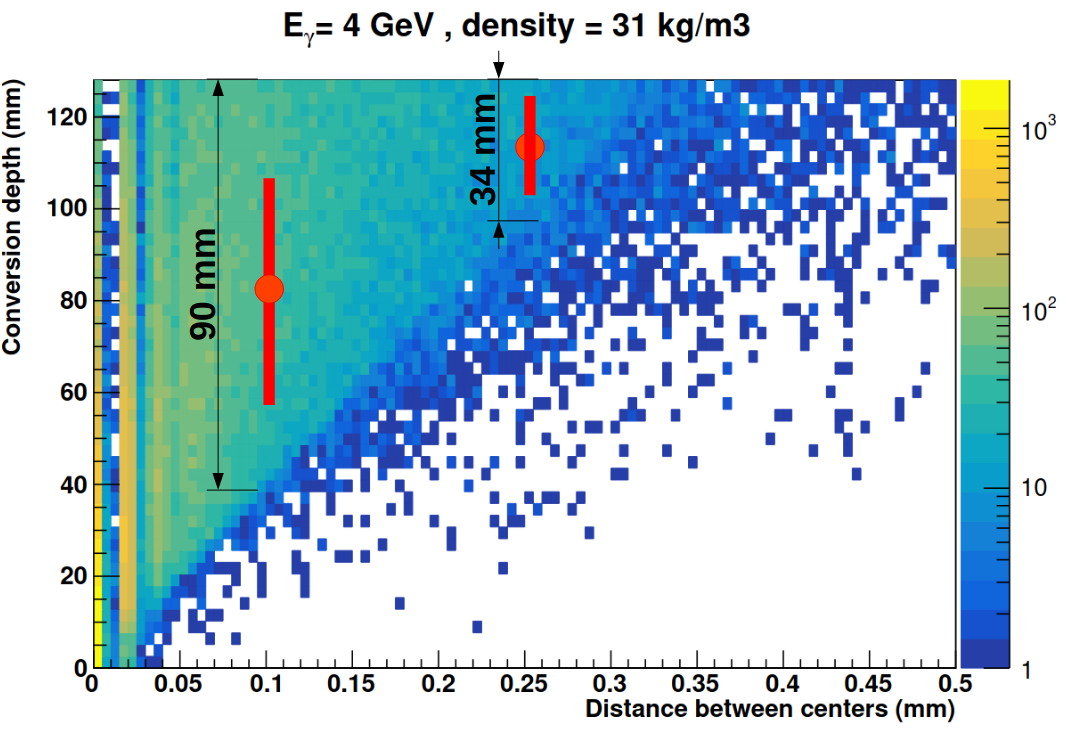}
  \caption{\label{fig:Radlength_desc} Left: For a cell length fixed at 80 mm, the converter length is modified and conversions are uniformly generated all along. To fix the radiation length, the density is inversely proportional to the converter length. Darker shades of green means a higher density. Right: 2D-histograms of the conversion depth as a function of the distance between the measurements in the first MAPS.}
\end{figure}

Although offering a high intrinsic analyzing power, the thin converter introduced in the previous studies has a low Figure-of-merit since its conversion efficiency is low. To increase the conversion, the radiation length of the converter must increase. However, as the radiation length increases , so does the multiple scattering spoiling the resolution on $\Phi_{ee}$. The goal is to find the perfect amount of material to maximize the Figure-of-merit. Not only scanning in radiation length, the actual length and density of the converter will be changed as well for a given radiation length as depicted in Figure~\ref{fig:Radlength_desc}. The minimal length of the cell is fixed at 80~mm, filling with air the cell when the converter does not fill it completely.\\

In this situation, the challenge is to \emph{``accurately''} derive the depth of the conversion and the associated uncertainty. Using the simulation, the distance between the measurements in the first MAPS is studied as a function of the conversion depth. Figure~\ref{fig:Radlength_desc} displays the bidimensional histogram representing the conversion depth as a function of the distance between the centers of position distributions on the first MAPS layer. A straightforward conclusion is that the conversion is most likely occuring deep in the converter if the distance between the two clusters in the MAPS is large. Consequently a minimal depth for the conversion can be inferred from the distance between the two leptons. As the converter has a finite size, the maximal depth is known as well. When the distance between the leptons is greater than 30$\mu$m, the distribution of conversion is almost uniform as seen by the projection for a 100~$\mu$m-distance in Figure~\ref{fig:Radlength_desc}. Therefore leptons are propagated in the middle of the minimal and maximal depth given the measured distance in the first MAPS, with an uncertainty being the available depth divided by $\sqrt{12}$. Two examples are illustrated in Figure~\ref{fig:Radlength_desc}, where red dots and bars represent the assumed conversion depth and the associated uncertainty for measurements distant by 100 and 250~$\mu$m on the first MAPS layer.

Below 30~$\mu$m, conversions occur mostly a few tens of millimeters upstream of the MAPS. Using the mean (and not the most likely) opening angle, the depth is calculated with a 100\%-uncertainty to account for the tail of conversions originating from deeper in the converter. 

\begin{figure}[!hbt]
  \centering
  \includegraphics[width=0.32\linewidth]{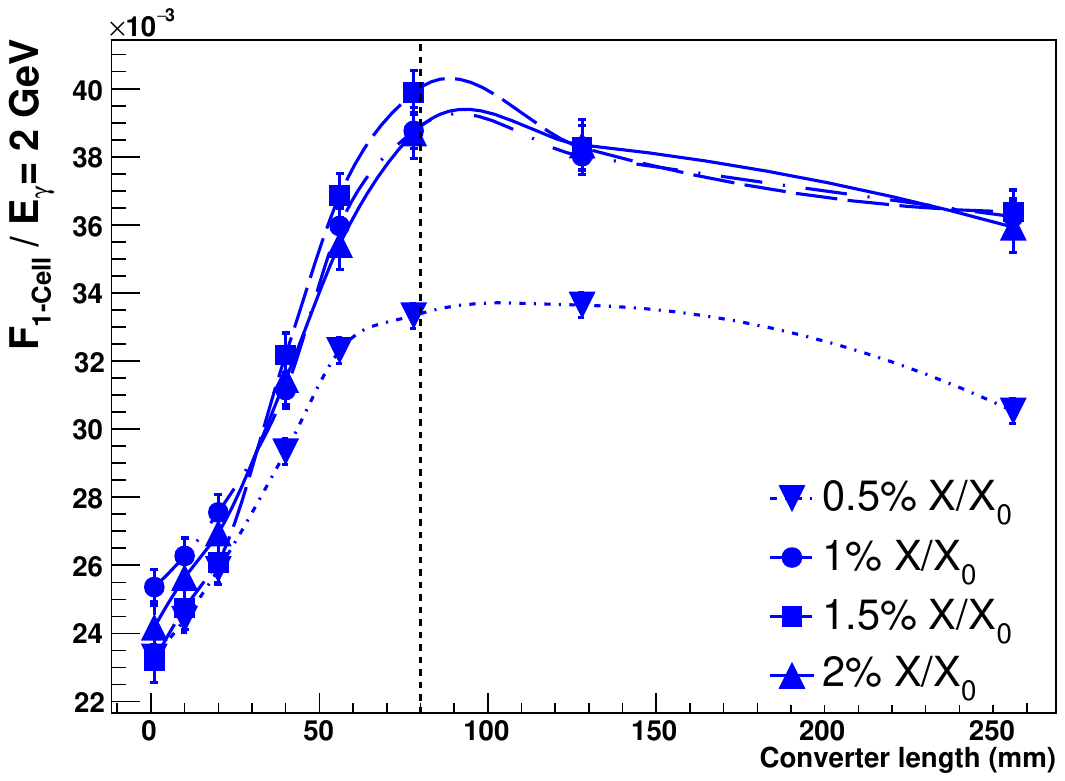} \includegraphics[width=0.32\linewidth]{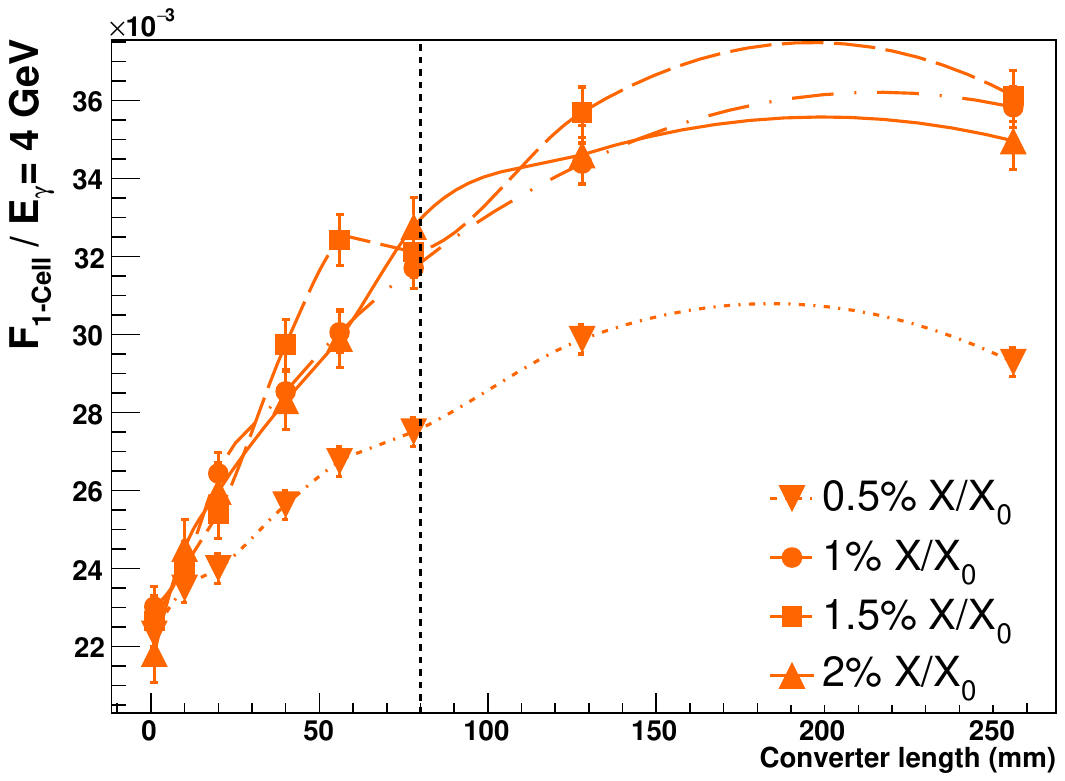}
  \includegraphics[width=0.32\linewidth]{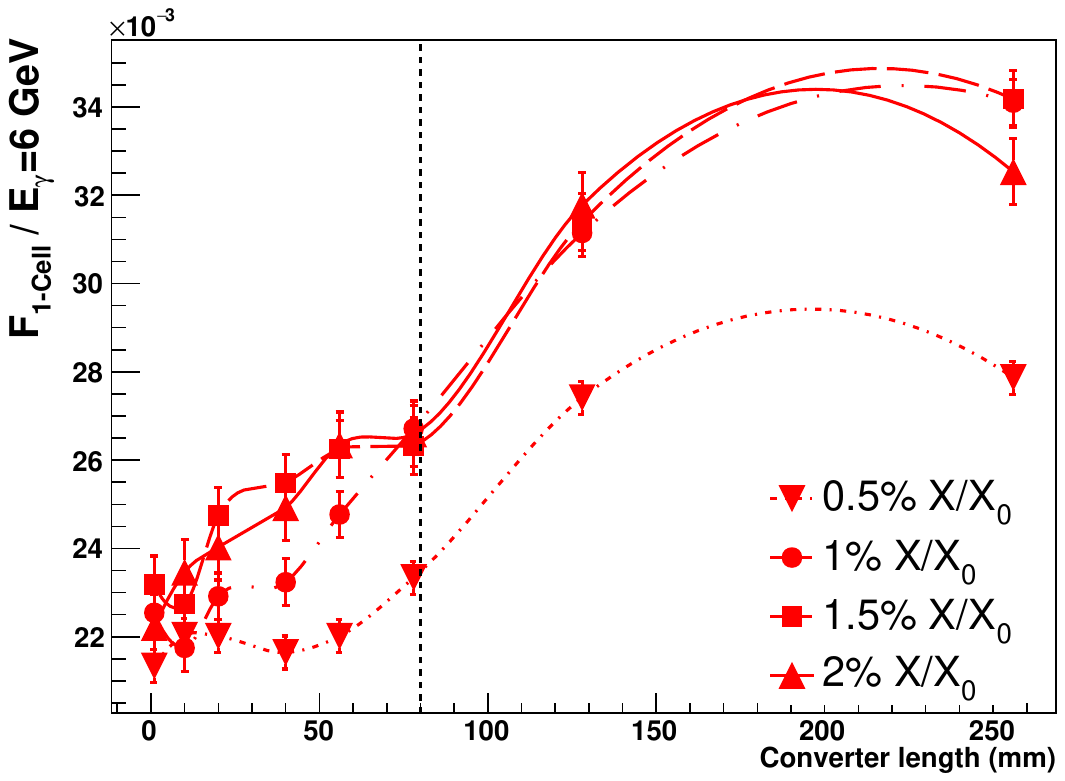}
  \caption{\label{fig:Radlength_res} Figure-of-merit studied as a function of the 1-cell converter radiation length for 2-GeV (top left), 4-GeV (top right) and 6-GeV (bottom) photons. For converter length smaller than 80~mm, a complementary layer of air is added to keep a 80-mm cell length.}
\end{figure}

Figure~\ref{fig:Radlength_res} represents the Figure-of-merit for a converter being 0.5, 1, 1.5 or 2\%~X/X0 with a length varying from 1 to 256~mm. The efficiency is calculated based on the conversion rate in a single cell. First the Figure-of-merit exhibits a strong dependence with the converter length and density at fixed radiation length. Whereas they are equal for a thin converter independently of the photon energy or radiation length, they increase by as much as 50\%-75\% with an extended converter. The lower the photon energy, the steeper the rise is and the sooner the maximum is reached. For 2~GeV, a maximum is seen around 80~mm, around 160~mm for 4~GeV and approximately 220~mm for 6~GeV. The behavior of the maximum of the FoM apparently proportional to the photon energy is a good indication about the reliability of the results. For 2~GeV-photons, a slow decrease of the FoM is observed when the converter gets longer than 80~mm.\\

The FoM with a 0.5\% X/X0-converter is much lower than with a converter of 1, 1.5 or 2\%~X/X0. For these three values, the FoM is almost identical but the efficiency compensates for the lower analyzing power obtained with 1.5 and 2\%~X/X0 compared to 1\%~X/X0. For an experimental measurement, a higher radiation length would increase by as much the rate in the last layers of MAPS. Consequently the design offering the highest FoM with the smallest radiation length is to be preferred. For a pair polarimeter offering the best performances from 2 to 6~GeV, the best cell is composed of 128~mm of Rohacell 31 commercially available.  

\subsection{Figure-of-merit for a 1m-long polarimeter}
\begin{figure}[!hbt]
  \centering
  \includegraphics[width=0.49\linewidth]{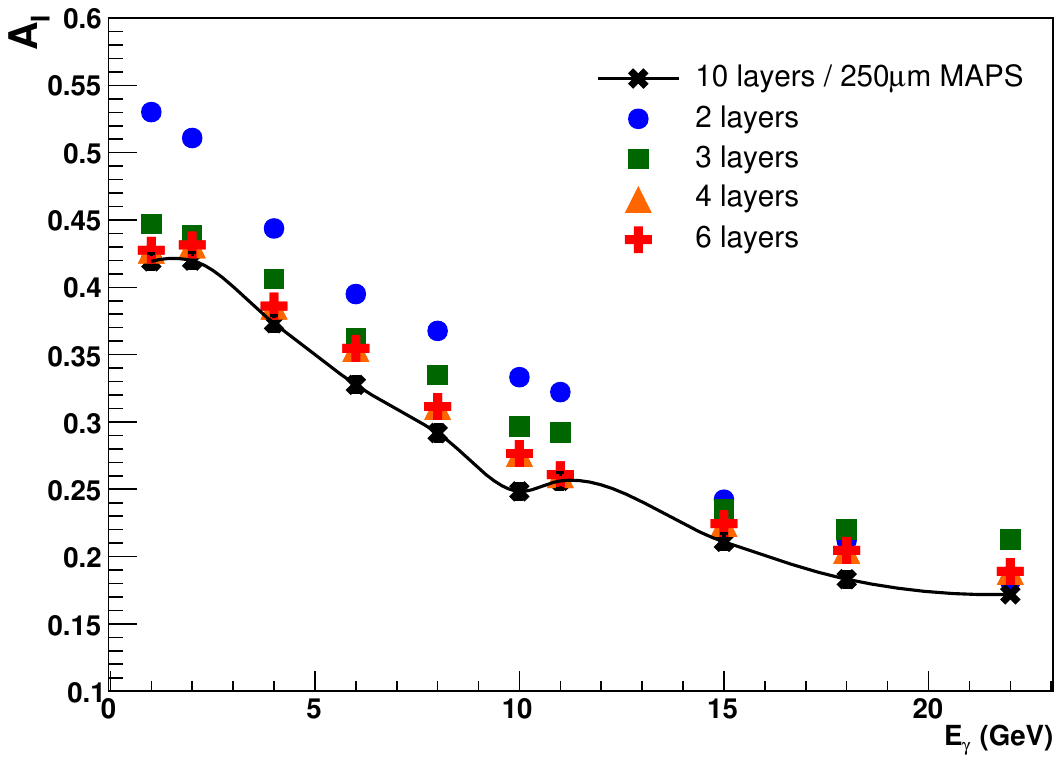}
  \includegraphics[width=0.49\linewidth]{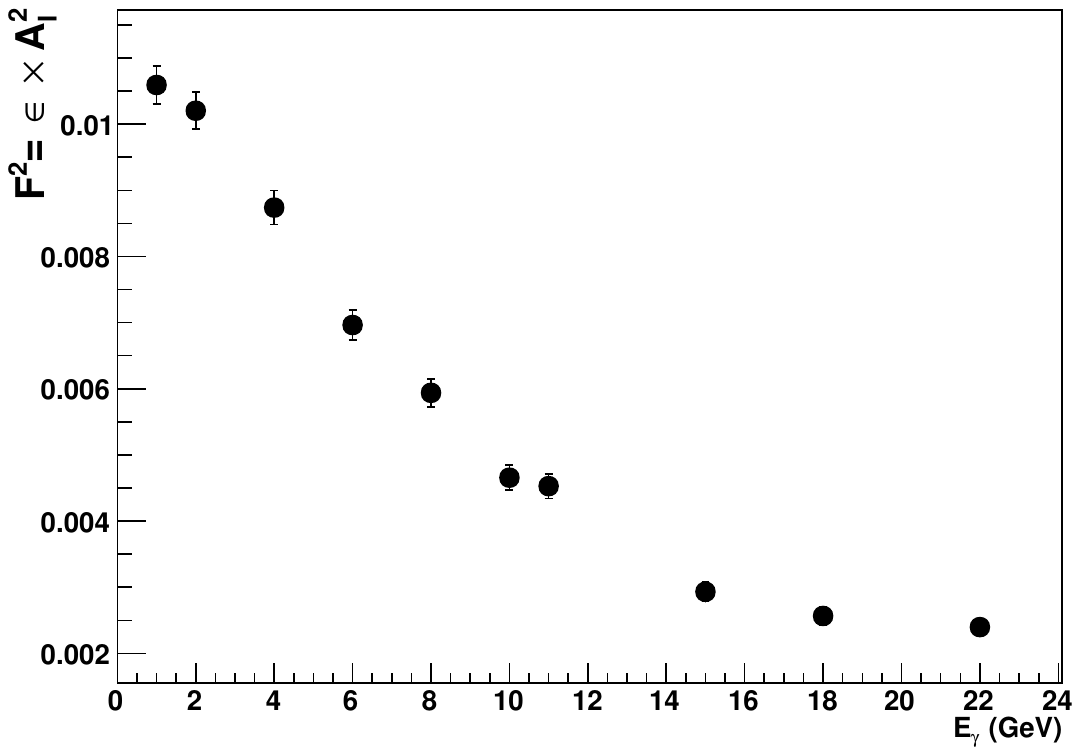}
  \caption{\label{fig:cell efficiency} Left: Intrinsic analyzing power as a function of the number of MAPS layers used to run the Kalman filter from 1 to 22~GeV. Right: Squared Figure-of-Merit for a 8-cell polarimeter with 128mm-long Rohacell31 converter.} 
\end{figure}

Assuming a maximal length of 1m, the polarimeter can be composed of 8 cells. In the previous studies, the Kalman filter was running over all 16 layers of MAPS for conversion occuring in the first cell. It is important to study the analyzing power as a function of the number of MAPS layer downstream of the conversion in order to estimate the polarimeter efficiency. Figure~\ref{fig:cell efficiency} displays the analyzing power as a function of the number of MAPS layers provided to the Kalman filter, for photon energies from 1 to 22~GeV.
The analyzing power is found stable for 3, 4 or 6 MAPS layers but gets surprisingly much higher with only 2 layers for photon energy below 10~GeV. This seems to indicate that the uncertainties in the covariance matrix may not be completely accurate with more than 3 layers. Using the standard multiple scattering description in the Kalman filter, it does underestimate the large angle part of Coulomb scattering . As a consequence the Kalman filter may underestimate the uncertainty of the predicted state vector. Consequently the choice between the direction-based and the position-based azimuthal angle may not be optimal. Not much can be done here to solve this reason except relaxing the covariance matrix arriving at the two last layers.
Still on Figure~\ref{fig:cell efficiency}, A$_I$ is shown with a MAPS thickness increased to 250$\mu$m to mimic ALPIDE sensors. The intrinsic analyzing power seems to be slightly lower than with a 0.05\%X/X0. Consequently ALPIDE sensors appear to suitable for a multi-GeV pair polarimeter. \\

Since only 2 layers of MAPS are enough, the total efficiency $\epsilon$ of this 1m-long polarimeter is given by:
\begin{equation}
  \epsilon=1-\exp\left(-\frac{7}{9} \times 0.01 \times (N_{_{cell}}-1)\right)=0.053\;,
\end{equation}
which is then combined to the 3-layer/50$\mu$m-MAPS analyzing power to get the total Figure-of-merit in Figure~\ref{fig:cell efficiency}.\\

To measure the linear polarization of a 2-GeV photon with an uncertainty $\sigma_{_{P}}=0.1$ with this pair polarimeter, a total number of one million photons would be required. However, for a 2-GeV photon, a 8cm-long cell with Rohacell71 (also commercially available) would allow to increase the number of cells from 8 to 12, increasing the efficiency from 5.3 to 8.2\% and reducing the number of required photons to 800 000. Accounting for the conversion in the MAPS which would represent a fifth of the total radiation length of a cell, this number would drop to 740 000.

\section{Conclusion}
In this article, we have introduced an innovative design for a pair polarimeter aimed at measuring the linear polarization of multi-GeV photons. Initially, monolithic active pixel sensors (MAPS) were considered to resolve the electron-positron pairs, given the sub-milliradian opening angles typical of high-energy photon conversions. To enhance conversion efficiency while mitigating the effects of multiple scattering, a low-density extended converter was subsequently proposed.

This optimized configuration yields a substantial improvement in performance: the intrinsic analyzing power of a single cell increases by 50-75\% for photon energies between 2 and 6~GeV. By comparing the Figure of Merit across various converter thicknesses, the lowest radiation-length converter emerges as the most favorable choice, as it effectively limits occupancy in the downstream MAPS layers within a multi-cell polarimeter.

The final choice of cell design will naturally depend on external factors, such as the photon energy spectrum and spatial constraints of the experimental environment. Nonetheless, for a device constrained to a total length of 1~meter, collecting just 750 000 photons at 2~GeV suffices to measure the linear polarization with a statistical precision of 0.1.

Thanks to this significant design optimization, precision measurements of the linear polarization of final-state photons in Deeply Virtual Compton Scattering (DVCS) are now most likely feasible at the Thomas Jefferson National Accelerator Facility. Moreover, this approach opens promising avenues for future applications in both astrophysics and fundamental field theory.

\section{Acknowledgements}
We warmly thank B.~Wojtsekhowski , P.~Laurent, L.~Chevalier and D.~Bernard for their continuous feedback and time spent explaining us all the subtleties of $\gamma$-polarimetry. We want to acknowledge the valuable support of M.~Jones, Hall leader at the Thomas Jefferson Laboratory, and all members of the NPS collaboration supporting and advertising this work. Finally this article would never have been written without my wife's, Marion Joly, unconditional love and support. 

\bibliographystyle{unsrt} 
\bibliography{biblio}

@article{bogdan,
	doi = {10.1117/1.jatis.4.1.011006},
  
	url = {https://doi.org/10.1117%2F1.jatis.4.1.011006},
  
	year = 2018,
	month = {mar},
  
	publisher = {{SPIE}-Intl Soc Optical Eng},
  
	volume = {4},
  
	number = {01},
  
	pages = {1},
  
	author = {Maxim Eingorn and Lakma Fernando and Branislav Vlahovic and Cosmin Ilie and Bogdan Wojtsekhowski and Guido Maria Urciuoli and Fulvio De Persio and Franco Meddi and Vladimir Nelyubin},
  
	title = {High-energy photon polarimeter for astrophysics},
  
	journal = {Journal of Astronomical Telescopes, Instruments, and Systems}
}

@article{de_Jager_2004,
	doi = {10.1140/epjad/s2004-03-045-5},
  
	url = {https://doi.org/10.1140%2Fepjad%2Fs2004-03-045-5},
  
	year = 2004,
	month = {feb},
  
	publisher = {Springer Science and Business Media {LLC}
},
  
	volume = {19},
  
	number = {S1},
  
	pages = {275--278},
  
	author = {C. de Jager and B. Wojtsekhowski and D. Tedeschi and B. Vlahovic and D. Abbott and J. Asai and G. Feldman and T. Hotta and M. Khadaker and H. Kohri and T. Matsumara and T. Mibe and T. Nakano and V. Nelyubin and G. Orielly and A. Rudge and P. Weilhammer and M. Wood and T. Yorita and R. Zegers},
  
	title = {A pair polarimeter for linearly polarized high energy photons},
  
	journal = {The European Physical Journal A}
}

@article{MAGER2016434,
title = {ALPIDE, the Monolithic Active Pixel Sensor for the ALICE ITS upgrade},
journal = {Nuclear Instruments and Methods in Physics Research Section A: Accelerators, Spectrometers, Detectors and Associated Equipment},
volume = {824},
pages = {434-438},
year = {2016},
note = {Frontier Detectors for Frontier Physics: Proceedings of the 13th Pisa Meeting on Advanced Detectors},
issn = {0168-9002},
doi = {https://doi.org/10.1016/j.nima.2015.09.057},
url = {https://www.sciencedirect.com/science/article/pii/S0168900215011122},
author = {M. Mager},
}

@misc{Birefringence,
      title={Probing vacuum birefringence under a high-intensity laser field with gamma-ray polarimetry at the GeV scale}, 
      author={Yoshihide Nakamiya and Kensuke Homma},
      year={2017},
      eprint={1512.00636},
      archivePrefix={arXiv},
      primaryClass={hep-ph},
      url={https://arxiv.org/abs/1512.00636}, 
}

@article{QuantumGravity,
   title={Lorentz violation at high energy: Concepts, phenomena, and astrophysical constraints},
   volume={321},
   ISSN={0003-4916},
   url={http://dx.doi.org/10.1016/j.aop.2005.06.004},
   DOI={10.1016/j.aop.2005.06.004},
   number={1},
   journal={Annals of Physics},
   publisher={Elsevier BV},
   author={Jacobson, Ted and Liberati, Stefano and Mattingly, David},
   year={2006},
   month=jan, pages={150–196}
   }

@article{ilie,
   title={Gamma-Ray Polarimetry: A New Window for the Nonthermal Universe},
   volume={131},
   ISSN={1538-3873},
   url={http://dx.doi.org/10.1088/1538-3873/ab2a3a},
   DOI={10.1088/1538-3873/ab2a3a},
   number={1005},
   journal={Publications of the Astronomical Society of the Pacific},
   publisher={IOP Publishing},
   author={Ilie, Cosmin},
   year={2019},
   month=sep, pages={111001} }

@article{HARPO,
title = {Performance measurement of HARPO: A time projection chamber as a gamma-ray telescope and polarimeter},
journal = {Astroparticle Physics},
volume = {97},
pages = {10-18},
year = {2018},
issn = {0927-6505},
doi = {https://doi.org/10.1016/j.astropartphys.2017.10.008},
url = {https://www.sciencedirect.com/science/article/pii/S0927650517301925},
author = {P. Gros \emph{et al.}}
}

@article{PDG,
    author = "Navas, S. and others",
    collaboration = "Particle Data Group",
    title = "{Review of particle physics}",
    doi = "10.1103/PhysRevD.110.030001",
    journal = "Phys. Rev. D",
    volume = "110",
    number = "3",
    pages = "030001",
    year = "2024"
}

@article{GEANT4,
title = {Geant4—a simulation toolkit},
journal = {Nuclear Instruments and Methods in Physics Research Section A: Accelerators, Spectrometers, Detectors and Associated Equipment},
volume = {506},
number = {3},
pages = {250-303},
year = {2003},
issn = {0168-9002},
doi = {https://doi.org/10.1016/S0168-9002(03)01368-8},
url = {https://www.sciencedirect.com/science/article/pii/S0168900203013688},
author = {S. Agostinelli \emph{et al.}},

}

\end{document}